\documentclass[fleqn,usenatbib]{mnras}

\usepackage[T1]{fontenc}

\DeclareRobustCommand{\VAN}[3]{#2}
\let\VANthebibliography\thebibliography
\def\thebibliography{\DeclareRobustCommand{\VAN}[3]{##3}\VANthebibliography}

\usepackage{graphicx}	% Including figure files
\usepackage{amsmath}	% Advanced maths commands
\usepackage{amssymb}	% Extra maths symbols
\usepackage{newtxtext,newtxmath}
\usepackage{multirow}
\usepackage{xcolor}
\usepackage{scalerel,tikz}
\usetikzlibrary{svg.path}
\definecolor{orcidlogocol}{HTML}{A6CE39}
\tikzset{orcidlogo/.pic={
 \fill[orcidlogocol] svg{M256,128c0,70.7-57.3,128-128,128C57.3,256,0,198.7,0,128C0,57.3,57.3,0,128,0C198.7,0,256,57.3,256,128z};
 \fill[white] svg{M86.3,186.2H70.9V79.1h15.4v48.4V186.2z}
 svg{M108.9,79.1h41.6c39.6,0,57,28.3,57,53.6c0,27.5-21.5,53.6-56.8,53.6h-41.8V79.1z M124.3,172.4h24.5c34.9,0,42.9-26.5,42.9-39.7c0-21.5-13.7-39.7-43.7-39.7h-23.7V172.4z}
 svg{M88.7,56.8c0,5.5-4.5,10.1-10.1,10.1c-5.6,0-10.1-4.6-10.1-10.1c0-5.6,4.5-10.1,10.1-10.1C84.2,46.7,88.7,51.3,88.7,56.8z};
}}
\newcommand\orcidicon[1]{\href{https://orcid.org/#1}{\mbox{\scalerel*{
\begin{tikzpicture}[yscale=-1,transform shape]
\pic{orcidlogo};
\end{tikzpicture}
}{|}}}}

\title[Role of turbulence driving for the IMF]{The role of the turbulence driving mode for the Initial Mass Function}

\author[Mathew, Federrath, \& Seta]{
Sajay Sunny Mathew$^{\orcidicon{0000-0002-8381-8195}\,1}$\thanks{E-mail: \href{mailto:sajay.mathew@anu.edu.au}{sajay.mathew@anu.edu.au}},
Christoph Federrath$^{\orcidicon{0000-0002-0706-2306}\,1,2}$\thanks{E-mail: \href{mailto:christoph.federrath@anu.edu.au}{christoph.federrath@anu.edu.au}}, and
Amit Seta$^{\orcidicon{0000-0001-9708-0286}\,1}$\thanks{E-mail: \href{mailto:amit.seta@anu.edu.au}{amit.seta@anu.edu.au}}
\\
$^{1}$Research School of Astronomy and Astrophysics, Australian National University, Canberra, ACT~2611, Australia\\
$^{2}$Australian Research Council Centre of Excellence in All Sky Astrophysics (ASTRO3D), Canberra, ACT~2611, Australia
}

\date{Accepted XXX. Received YYY; in original form ZZZ}

\pubyear{2022}

\begin{document}
\label{firstpage}
\pagerange{\pageref{firstpage}--\pageref{lastpage}}
\maketitle

\begin{abstract}
Turbulence is a critical ingredient for star formation, yet its role for the initial mass function (IMF) is not fully understood. Here we perform magnetohydrodynamical (MHD) simulations of star cluster formation including gravity, turbulence, magnetic fields, stellar heating and outflow feedback to study the influence of the mode of turbulence driving on IMF. We find that simulations that employ purely compressive turbulence driving (COMP) produce a higher fraction of low-mass stars as compared to simulations that use purely solenoidal driving (SOL). The characteristic (median) mass of the sink particle (protostellar) distribution for COMP is shifted to lower masses by a factor of $\sim 1.5$ compared to SOL. Our simulation IMFs capture the important features of the observed IMF form. We find that turbulence-regulated theories of the IMF match our simulation IMFs reasonably well in the high-mass and low-mass range, but underestimate the number of very low-mass stars, which form towards the later stages of our simulations and stop accreting due to dynamical interactions. Our simulations show that for both COMP and SOL, the multiplicity fraction is an increasing function of the primary mass, although the multiplicity fraction in COMP is higher than that of SOL for any primary mass range. We find that binary mass ratio distribution is independent of the turbulence driving mode. The average specific angular momentum of the sink particles in SOL is a factor of 2 higher than that for COMP. Overall, we conclude that the turbulence driving mode plays a significant role in shaping the IMF.
\end{abstract}


%%%%%%%%%%%%%%%%%%%%%%%%%%%%%%%%%%%%%%%%%%%%%%%%%%

%%%%%%%%%%%%%%%%% BODY OF PAPER %%%%%%%%%%%%%%%%%%

\section{Introduction}
Supersonic turbulence pervades the interstellar medium (ISM) and it is a critical ingredient for star formation in molecular clouds (MC). Turbulence, by moving the gas around randomly, counteracts a monolithic collapse of the cloud driven by gravity and serves as a primary agent for the low star formation rate observed in the Milky Way and nearby galaxies \citep{2005ApJ...630..250K,2009ApJS..181..321E,2011ApJ...729..133M,2012ApJ...761..156F,2015MNRAS.450.4035F,2016ApJ...831...73V,2016ApJ...833..229L}. On the other hand, it also creates localised compressions within the clouds, enhancing the growth of high-density regions, which are potential sites of star formation. Thus, turbulence plays a fundamental role in regulating star formation. Numerical studies establish that the gas density probability distribution function (PDF) of supersonic turbulence is approximately log-normal \citep{1994ApJ...423..681V,1997MNRAS.288..145P,2007ApJ...665..416K,2008ApJ...688L..79F,2013MNRAS.436.1245F,2013MNRAS.430.1880H,2015MNRAS.448.3297F,2022MNRAS.514..957S}. The density statistics of turbulence, particularly the log-normal nature of the gas density PDF, along with the velocity statistics have been used to explain the observed star formation efficiency \citep{2013ApJ...763...51F}, star formation rate \citep{2005ApJ...630..250K,2009ApJ...699..850K,2011ApJ...743L..29H,2012ApJ...761..156F,2015MNRAS.450.4035F} and the initial mass function (IMF) \citep{2002ApJ...576..870P,2008ApJ...684..395H,2009ApJ...702.1428H,2012MNRAS.423.2037H,2013MNRAS.430.1653H}.

The IMF refers to the mass distribution of young stars, e.g., in young star clusters, and it serves as the PDF for the mass of a star when it reaches the main sequence phase. The form of the IMF is found to be remarkably similar in different star-forming regions in the local neighbourhood and beyond, i.e., it is thought to be relatively universal \citep[see the reviews by][]{2010ARA&A..48..339B,2014prpl.conf...53O,2018PASA...35...39H,2020SSRv..216...70L}, although there are studies that suggest that the IMF might also differ \citep[e.g.,][]{2014MNRAS.444.1957D,2017MNRAS.464.1738D}. The IMF is a power law at high masses and the number of stars $N(M)$ can be defined via the relation, $dN \propto M^{-1.35}\, d\mathrm{log}M\, (M > 1\, \mathrm{M_\odot})$ \citep{1955ApJ...121..161S}. The IMF flattens at lower masses and the mass distribution of the sub-solar range can be represented by a segmented power law \citep{2001MNRAS.322..231K} or a log-normal function \citep{2005ASSL..327...41C}. The peak mass or the characteristic mass of the IMF is located at around $0.2$--$0.3\, \mathrm{M_\odot}$ \citep{2003PASP..115..763C,2008ApJ...681..365E,2014prpl.conf...53O}.

The standard deviation of the turbulent gas density PDF ($\sigma_\rho$) is proportional to the rms Mach number of the gas flow ($\mathcal{M}$) and is given by $\sigma_\rho/\langle \rho \rangle=b \mathcal{M}$, where $\langle \rho \rangle$ is the mean density. The value of the proportionality constant $b$ is dependent on the mode of turbulence driving \citep{1997MNRAS.288..145P,1998PhRvE..58.4501P,2008ApJ...688L..79F}. Purely compressive (curl-free) driving corresponds to $b\sim1$ and purely solenoidal (divergence-free) driving corresponds to a value of $b\sim1/3$ \citep{2008ApJ...688L..79F,2010A&A...512A..81F}. Values between $1$ and $1/3$ represent a mixture of compressive and solenoidal modes. Hence, the width of the gas density PDF is a function of the relative importance of the two driving modes. Dynamical mechanisms (such as galactic spiral shocks, and accretion) as well as supernova explosions and other stellar feedback mechanisms like radiation-pressure-driven shells tend to induce more compressive (curl-free) modes of turbulence in MCs, whereas shear and magneto-rotational instability excite more solenoidal (divergence-free) modes \citep{2017IAUS..322..123F}. The prospective influence on the IMF as a result of the dependence of the gas density PDF on the turbulence driving mode has been studied in a few numerical works \citep{2010A&A...516A..25S,2011MNRAS.413.2741G,2015MNRAS.449..662L,2017MNRAS.465..105L}, although a continuous turbulence driving was not employed in most of these studies, which is crucial to establish fully-developed turbulence statistics.

Here we investigate the effect of the mode of turbulence driving in setting the IMF. In addition to gravity and turbulence, we also include other important physics for the IMF: magnetic fields, protostellar heating and outflow feedback \citep{2020MNRAS.496.5201M,2021MNRAS.507.2448M}. We perform multiple simulations with different turbulence realisations, such that we have a statistically meaningful sample to study the IMF. We also study how the stellar multiplicity properties are influenced by the mode of turbulence driving.

In Section~\ref{sec:method}, we describe the numerical methodology and turbulence setup, in particular the turbulence driving method that allows us to control the relative fraction of solenoidal and compressive modes in the driving field. We also explain the sub-grid models for stellar radiative heating and jets/outflows, and provide the initial conditions for the simulations. In Section~\ref{sec:results}, we study the influence of the turbulence driving mode in the star cluster formation process by comparing the results of simulations that employ a purely compressive mode of driving with simulations characterised by a purely solenoidal mode of driving. For each of the two models, we investigate the column density and temperature structure, evolution of dynamical quantities, and the mass distribution of the stars formed in our simulations. In Section~\ref{sec:IMF_comp}, we compare the protostellar mass distribution from our simulations with the IMF derived from observations and theoretical models. We examine the multiplicity and the stellar angular momentum in Section~\ref{sec:Multi}. In Section~\ref{sec:discussions}, we discuss some of the previous numerical works on the influence of turbulence on the IMF. The primary results and conclusions are discussed in Section~\ref{sec:conclude}.

\section{Methods}
\label{sec:method}
\subsection{Basic numerical methods and magnetohydrodynamics}
To perform the numerical modelling of star cluster formation, we solve the magnetohydrodynamical (MHD) equations with gravity on an adaptive mesh refinement (AMR) \citep{1989JCoPh..82...64B} grid, using the PARAMESH library \citep{MACNEICE2000330} in a significantly modified version of the \textsc{flash} (version~4) code \citep{2000ApJS..131..273F,2008ASPC..385..145D},
\begin{equation}
\frac{\partial\rho}{\partial t}+\nabla\cdot (\rho \mathbf v) = 0,
\end{equation}
\begin{equation}
\left(\frac{\partial}{\partial t} + \mathbf v \cdot \nabla \right)\, \mathbf v = \frac{(\mathbf B \cdot \nabla) \mathbf B}{4 \pi \rho} - \frac{\nabla P_{\mathrm{tot}}}{\rho} + \mathbf g + \mathrm{\mathbf{f}_{stir}}, \label{eq:mhd2}
\end{equation}
\begin{equation} 
\frac{\partial \mathbf B} {\partial t} = \nabla \times (\mathbf v \times \mathbf B),   \hspace{4mm}\nabla \cdot \mathbf B = 0,
\end{equation}
where $\rho,\mathbf v, \mathbf B, P_{\mathrm{tot}} = P + 1/(8\pi) |\mathbf B |^2, $ and $\mathrm{\mathbf{f}_{stir}}$ correspond to the gas density, velocity, magnetic field, pressure (sum of thermal and magnetic pressures), and turbulent acceleration field, respectively. Here $\mathbf{g}$ is the gravitational acceleration and is the sum of the self-gravity of the gas and the acceleration as a result of the mass of sink (star) particles (see \S\ref{sec:sink}). We utilise the 5-wave HLL5R approximate Riemann method to solve the MHD equations \citep{2011JCoPh.230.3331W}. The self-gravity of the gas is evaluated using a multi-grid Poisson solver \citep{2008ApJS..176..293R}.

\subsection{Turbulence driving}
\label{sec:turb}
We drive turbulent motions in our simulations through the specific forcing term $\mathrm{\mathbf{f}_{stir}}$ in the MHD equations (see Eq.~\ref{eq:mhd2}). The acceleration field $\mathrm{\mathbf{f}_{stir}}$ is modelled using a stochastic Ornstein-Uhlenbeck (OU) process \citep{1988CF.....16..257E,schmidt,2010A&A...512A..81F}. The OU process enables us to continuously drive turbulence with an $\mathrm{\mathbf{f}_{stir}}$ field that varies smoothly in space and time. If no Helmholtz decomposition is carried out, the output of such a process is a natural mixture of stirring modes, i.e., a 2:1 mixture of solenoidal ($\nabla \cdot \mathbf{f_{\mathrm{stir}}} = 0$) to compressive ($\nabla \times \mathbf{f_{\mathrm{stir}}} = 0$) modes. Using the respective projection in Fourier ($k$) space, we can decompose the acceleration field obtained from the OU process into purely solenoidal and purely compressive components, and depending on the requirement, we can choose to drive turbulence with any one of these components or with a mixture of the two. The projection operator in $k$-space is given by \citep{2008ApJ...688L..79F}
\begin{equation}
    \mathcal{P}_{ij}^\zeta (\mathbf{k}) = \zeta \mathcal{P}_{ij}^\perp (\mathbf{k}) + (1-\zeta) \mathcal{P}_{ij}^\parallel (\mathbf{k}) = \zeta \delta_{ij} + (1-2\zeta) \frac{k_i k_j}{\lvert k \rvert^2}.
\end{equation}
The value of $\zeta$ controls the relative strength of solenoidal and compressive modes. By setting $\zeta=1$, we can obtain the solenoidal component of the acceleration field, while $\zeta=0$ gives the compressive component. We refer the reader to \citet{2008ApJ...688L..79F,2010A&A...512A..81F} for a more detailed description of the OU process associated with the turbulence driving method used here.

Our forcing module is configured to inject kinetic energy only on the largest scales (wave numbers $k=1\dots 3$, where $k$ is in units of $2\pi/L$ with the side length $L$ of the box) by using a parabolic function for the amplitude with the peak at $|\mathbf{k}|=2$ and zero amplitude at $|\mathbf{k}|=1,3$. Such a treatment allows the injected kinetic energy to naturally cascade down to smaller scales, resulting in a velocity power spectrum $\sim k^{-2}$ or equivalently a velocity dispersion -- size relation of $\sigma_v \propto \ell^{1/2}$, as we set the overall amplitude such that the turbulence has a sonic Mach number of $\mathcal{M}=5$, a typical configuration for molecular clouds \citep{1981MNRAS.194..809L,2002A&A...390..307O,2004ApJ...615L..45H,2011ApJ...740..120R,2013MNRAS.436.1245F,2021NatAs...5..365F}. The turbulence driving module used here is publicly available \citep{2022ascl.soft04001F}.

\subsection{Star formation (sink particles) and AMR}
\label{sec:sink}
Sink particles are used for modelling the collapsing, high-density regions of a cloud. When the density of the central part of a collapsing core becomes too high to resolve and the associated time-scale becomes too small to follow with AMR, the gravitational bound gas in the inner regions is replaced by a sink particle. To prevent artificial sink particle formation, in addition to the requirement that the gas constituting a sink particle be gravitationally bound, we carry out a suite of tests as implemented by \citet{2010ApJ...713..269F} before transforming gas to sink particles locally. The sink particles are introduced in a spherical control volume described by a given radius (here equal to the accretion radius of the sink particle) and centred at the cell at which the density is higher than the threshold density which in turn is decided by the Jeans length,
\begin{equation}
    \rho_{\mathrm{sink}} = \frac{\pi\, c_\mathrm{s}^2}{G\, \mathrm{\lambda_J^2}} = \frac{\pi\, c_\mathrm{s}^2}{4\,G\, r_{\mathrm{sink}}^2},
\end{equation}
where $c_\mathrm{s}$ is the sound speed, $G$ is the gravitational constant, $\mathrm{\lambda_J}=[\pi c_s^2/(G\rho)]^{1/2}$ is the local Jeans length, and $r_{\mathrm{sink}}= \lambda_\mathrm{J}/2$ is the sink particle radius.

In order to be conforming with the \citet{1997ApJ...489L.179T} criterion to avoid fragmentation artificially, the radius $r_{\mathrm{sink}}$ of the sink particle is set such that  $2r_{\mathrm{sink}}=5\, \Delta x$, where $\Delta x$ is the size of the grid cell on the highest level of refinement. On all lower AMR levels, $\lambda_\mathrm{J}$ is always resolved with a minimum of 16 grid cell lengths to ensure that the turbulent flow is reasonably well resolved on the scales of a Jeans length \citep{2011ApJ...731...62F}.

At every accretion step, the mass, linear momentum and angular momentum of each sink particle are updated by following the conservation laws. The new position of the sink particle after accretion is determined by the centre of mass of the sink particle and the accreted material. An intrinsic angular momentum (spin) is assigned to the sink particle, which stores the accreted angular momentum, ensuring the conservation of the total angular momentum. The rotational axis of the sink particle along which jets and outflows are launched is determined by the spin \citep[][]{2014ApJ...790..128F}; see further details in Sec.~\ref{sec:feedback}.

All gravitational interactions of the sink particles between each other and with the gas are computed by direct summation over all the sink particles and grid cells \citep{2011IAUS..270..425F}. A second-order leapfrog integrator is utilised to advance the sink particles in time.

\subsection{Equation of state (EOS)}
\label{sec:polytropic}
The temperature structure of the gas in dense cores is controlled by a combination of different thermodynamical mechanisms including cosmic-ray heating, compressional heating, and cooling by dust grains \citep{1973FCPh....1....1L,1998ApJ...495..346M}. The initial phase of the collapse is approximately isothermal while the cores are still optically thin \citep{1995ApJ...443..152W,2000ApJ...531..350M,2010MNRAS.404....2G}. However, as the density in the central regions increases, the gravitational energy is not readily radiated away and the temperature of the core starts to increase due to compressional heating. Thus, the collapse transitions from an isothermal to an adiabatic process. To accurately model the thermal evolution of the gas, the equation of energy conservation has to be solved simultaneously with the radiation transfer (RT) equation. Solving the RT equation involving every grid cell and for every timestep is computationally demanding \citep{2022MNRAS.512..401M}, and thus incorporating it in these large-scale simulations is a challenging task. In order to enable a large statistical study, instead of solving the RT equations, we use an approximation, by closing the system of MHD equations with a polytropic equation of state for the gas pressure $P=P_\mathrm{EOS}$, given by
\begin{equation}
    P_\mathrm{{EOS}} = c_\mathrm{s}^2\, \rho^{\gamma}.
\end{equation}
Utilising the ideal gas EOS, the corresponding temperature is derived as
\begin{equation}
    T_{\mathrm{EOS}} =  \frac{\mu\, m_{\mathrm{H}}}{k_\mathrm{B}\, \rho}\, P_\mathrm{{EOS}} = \frac{\mu\, m_{\mathrm{H}}}{k_\mathrm{B}}\, c_\mathrm{s}^2\, \rho^{\gamma-1}\,.
\end{equation}
Here $c_\mathrm{s}^2=(0.2\,\mathrm{km/s})^2$ is the square of the sound speed in the isothermal range ($\gamma = 1$) for solar metallicity, molecular gas at $10\,\mathrm{K}$, and $\mu = 2.35$ is the mean molecular weight (in units of the atomic mass of hydrogen, $m_{\mathrm{H}}$). The polytropic exponent is then adjusted based on the local density of the gas, as
\begin{equation}
    \gamma =
      \begin{cases}
        1   & \text{for \hspace{7mm} $\rho \le \rho_1 \equiv 2.50 \times 10^{-16}\, \mathrm{g\, cm^{-3}}$,}\\ 
        1.1 & \text{for\, $\rho_1 < \rho \le \rho_2 \equiv 3.84 \times 10^{-13}\, \mathrm{g\, cm^{-3}}$,}\\ 1.4 & \text{for\, $\rho_2 < \rho \le \rho_3 \equiv 3.84 \times 10^{-8}\, \mathrm{g\, cm^{-3}}$,}\\
        1.1 & \text{for\, $\rho_3 < \rho \le \rho_4 \equiv 3.84 \times 10^{-3}\, \mathrm{g\, cm^{-3}}$,}\\
        5/3 & \text{for  \hspace{7mm} $\rho > \rho_4$.}
      \end{cases}
\end{equation} 
The value of the polytropic exponent $\gamma$ changes with the local gas density, and is based on previous detailed radiation-hydrodynamic simulations of the formation of protostars. It covers the isothermal phase during the initial collapse, adiabatic heating during the formation of the first and second core, and the influence of $\mathrm{H_2}$ dissociation during the second collapse \citep{1969MNRAS.145..271L,1993ApJ...411..274Y,2000ApJ...531..350M,2009ApJ...703..131O}. However, it does not consider the increase in thermal gas pressure due to the stellar radiative heating (feedback), which is discussed next.

\subsection{Stellar feedback} \label{sec:feedback}
\subsubsection{Radiative heating}
\label{sec:radfeedback}
Stars in their early stages of formation have high accretion luminosities, which can suppress fragmentation, enabling the existing stars to reach high masses by continued accretion \citep{2009MNRAS.392.1363B,2011ApJ...740...74K,2016MNRAS.458..673G,2017JPhCS.837a2007F,2020MNRAS.496.5201M,2020arXiv201003539H}. Thus, it is crucial to take into consideration the temperature variation due to the stellar heating feedback. To precisely model the stellar heating, the RT equation has to be solved together with the energy conservation equation, as mentioned in Sec.~\ref{sec:polytropic}, which involves tracing the rays emitted from the protostars and the rays absorbed or scattered by dust grains. Solving the RT equation in large-scale simulations is extremely challenging because of the computational expense \citep{2016NewA...43...49B,2022MNRAS.512..401M}. As an alternative, we will employ the polar heating model developed by \citet{2020MNRAS.496.5201M} to model the direct stellar heating. The polar heating model is based on the heating model in \citet{2017JPhCS.837a2007F} and takes into account the shielding of the radiation field by the dust particles in the accretion disc. Following the works of \citet{2004A&A...417..793P} and \citet{2016NewA...43...49B}, our model assumes a disc density distribution around each sink particle (protostar) that is determined by the radial distance $r$ and the angle $\theta$ subtended from the angular momentum axis of the sink particle. The stellar radiant power is distributed over the grid cells surrounding the sink particle based on this dust/disc density distribution. 

The radiation from the central star is absorbed by the dust particles with the rate of energy absorption given by
\begin{equation}
    Q (r, \theta) = \chi\, \frac{L_{\star}}{4\pi r^2}\, \exp\left(-\tau (r,\theta)\right), \label{eq:Qheat}
\end{equation}
where $\chi$ is the absorption coefficient. The star's luminosity ($L_{\star}$), which consists of both the accretion and intrinsic luminosities, is estimated by employing the protostellar evolution model by \citet{2009ApJ...703..131O}. The total optical depth ($\tau$) in any direction given by $\theta$ is
\begin{equation}
    \tau = \int \kappa\, \rho(r,\theta)\, \mathrm{d}r, 
\end{equation}
where $\kappa$ is the grey opacity (a constant here) and $\rho(r,\theta)$ corresponds to the dust/disc density distribution assumed \citep[see][for a detailed discussion of the analytical model of the disc density distribution employed here]{2020MNRAS.496.5201M}. The dust grains in the disc can absorb the radiation and therefore the field will be diminished in the directions of the disc, and the primary heating will be restricted to the polar directions.

The dust grains will achieve an equilibrium temperature when they emit the same amount of energy they absorb. Thus, we can write
\begin{equation}
    \frac{\sigma_\mathrm{{SB}}}{\pi}\, \chi\, T_{\mathrm{heat}}^4 = \frac{Q}{4\pi},
\end{equation}
where $\sigma_\mathrm{{SB}}$ is the Stefan-Boltzmann constant and $T_{\mathrm{heat}}$ is the temperature due to stellar heating. We note that the model ignores the reprocessed radiation field, but the change in temperature due to the presence of the reprocessed field is minimal and would barely affect the IMF.

The space-dependent pressure term derived from the polar stellar heating module is added to the pressure calculated from the polytropic equation of state to accomodate the change in temperature or equivalently the change in thermal pressure due to the stellar radiative heating  \citep[see][]{2016MNRAS.458..673G,2018AAS...23111403G,2017JPhCS.837a2007F}. Thus, the final gas pressure is
 \begin{align}
    P &= \left[P^4_{\mathrm{EOS}} + P^4_{\mathrm{heat}}\right]^{1/4} \nonumber \\ 
      &= \left[P^4_{\mathrm{EOS}} + {\left(\frac{k_\mathrm{B}\, \rho}{\mu\, m_\mathrm{H}}\right)}^4\, T^4_{\mathrm{heat}}\right]^{1/4}, \label{eq:Pheat}
\end{align}
which is used in the MHD momentum equation, Eq.~(\ref{eq:mhd2}).

\subsubsection{Jets/Outflows}
\label{sec:outflowfeedback}
The bipolar mechanical feedback from protostars consists of jets which are highly collimated fast streams of gas that penetrate through the accreting envelope, and the wide-angle low-speed molecular outflows \citep{2014prpl.conf..451F}. All young stars lose part of their mass through jets and mass outflows \citep{2000prpl.conf..867R,2002ApJ...580..336W}. The material ejected from stars or young stellar objects (YSOs) also disperses the gas envelope surrounding the protostar, creating cavities. The gap in the mass scale between the core mass function (CMF) and the IMF is generally considered to be caused by the mass loss in protostars as a result of jets and outflows, which is often parameterized by a mass-independent core-to-star efficiency $\epsilon \sim 0.25 - 0.5$ \citep{2000ApJ...545..364M,2008ApJ...687..340M,2012ApJ...761..156F,2014ApJ...790..128F,2014ApJ...784...61O}. Two primary effects of the inclusion of jets/outflows in simulations are the reduction in the star formation rate and the increase in the number of protostellar objects formed \citep{2014ApJ...790..128F,2021MNRAS.502.3646G,2021MNRAS.507.2448M}. Thus, the incorporation of outflow feedback in numerical works is essential to produce conclusive results on the IMF.

We include jet/outflow feedback in our simulations by using the subgrid-scale (SGS) outflow model developed by \citet{2014ApJ...790..128F}. It captures both the low-speed molecular outflows and the fast jet components and includes angular momentum transfer. The SGS module redistributes momentum among the grid cells enclosed within a control volume determined by two conical sections about the sink particle. The conical sections open towards the opposite poles of the sink particle and are defined by an opening angle $\theta_{\mathrm{out}}=30^{\circ}$ \citep{1982MNRAS.199..883B} measured from the angular momentum axis. We fix the radial extent (height of the cone) equal to $r_{\mathrm{out}}=16\Delta x$ measured from the sink particle's position (tip of the cone), where $\Delta x$ is the cell size on the highest AMR level, as done in \citet{2014ApJ...790..128F}, to ensure convergence. Radial and angular smoothing kernels are used to attain smooth transition at the interface. The momentum injected into each of the cones is
\begin{equation}
    \mathbf{P_{\mathrm{out}}}=\pm (1/2)\, M_{\mathrm{out}}\, \mathbf{V_{\mathrm{out}}},
\end{equation}
where $M_{\mathrm{out}}$ corresponds to the mass ejected, which is equivalent to the fraction $f_{\mathrm{m}}$ of the mass accreted by the sink particle in a timestep $\Delta t$, i.e., $M_{\mathrm{out}} = f_{\mathrm{m}}\, \dot M_{\mathrm{acc}}\, \Delta t$. We define $f_{\mathrm{m}}=0.3$ \citep{2014ApJ...790..128F}, which agrees with observational surveys \citep{1995AJ....109.1846H,2007A&A...468L..29C,2011ApJ...737L..26B}, theoretical models of the outflow feedback \citep{1982MNRAS.199..883B,1988ApJ...328L..19S,2007prpl.conf..277P}, and the estimates from other numerical simulations \citep{2008A&A...477....9H,2012MNRAS.422..347S,2013ApJ...774...12F}.

$\mathbf{V_{\mathrm{out}}}$ is set to the Kepler speed close to the protostellar surface, such that
\begin{equation}
    |\mathbf{V_{\mathrm{out}}}| = 100\, \mathrm{km\, s^{-1}} \left(\frac{M_{\mathrm{sink}}}{0.5\, \mathrm{M_\odot}}\right)^{1/2},
\end{equation}
where $M_{\mathrm{sink}}$ is the sink particle mass and $100\, \mathrm{km\, s^{-1}}$ is the typical jet speed (and Kepler speed) for a protostar of mass $M \sim 0.5\, \mathrm{M_\odot}$ at a radius of $R \sim 10\, \mathrm{R_\odot}$. $\mathbf{V_{\mathrm{out}}}$ consists of a slow component with a speed of $0.25\, |\mathbf{V_{\mathrm{out}}}|$ and a high-speed component with a speed of $0.75\, |\mathbf{V_{\mathrm{out}}}|$. The momentum injection in the cones associated with the fast component is limited to an opening angle of $5^{\circ}$. Utilising such a velocity profile ensures that the faster jet and the slower molecular outflow components are distinguished.

The model removes a fraction $f_\mathrm{a}$ of the angular momentum accreted by the sink particle and re-introduces it to the jet and outflow components. We employ the default value of $f_\mathrm{a}=0.9$ in the SGS model, which is based on the observations in \citet{2002ApJ...576..222B} and previous numerical studies \citep[e.g.][]{2006ApJ...641..949B,2008A&A...477....9H}. 

The MHD code self-consistently carries away the momentum inserted into the two cones to larger distances. Through a series of rigorous tests, \citet{2014ApJ...790..128F} have shown that the large-scale outflow features, that is, the mass, linear momentum, angular momentum, and outflow speed, converge independent of the resolution with the SGS outflow model. We refer the reader to \citet{2014ApJ...790..128F} and references therein for more details of the SGS model and justification of the parameter choices. 
% =================================
\subsection{Initial conditions and simulation parameters}
\label{sec:parameters}

The simulations are performed in a three-dimensional triple-periodic computational box with side length $L=2\, \mathrm{pc}$. At the highest level of refinement, we allow for a maximum effective grid resolution of $N_{\mathrm{eff,\, res}}^3=4096^3$ cells or a minimum cell size of $\Delta x_{\mathrm{cell}}=100\,\mathrm{AU}$. The initial gas density is uniform with $\rho_{\circ} = 6.56 \times 10^{-21}\, \mathrm{g\, cm^{-3}}$, which yields a total cloud mass of $M_{\mathrm{cl}}=775\, \mathrm{M_{\odot}}$  and a mean free-fall time of $t_{\mathrm{ff}}= 0.82\,$Myr. Initially, the turbulence driving module stirs the gas in the computational domain in the absence of self-gravity. To ensure that a fully-developed turbulent state is reached, self-gravity is activated only after two turbulent crossing times, $2 t_\mathrm{turb}=L/(\mathcal{M}c_\mathrm{s}) = 2\,\mathrm{Myr}$ \citep{2010A&A...512A..81F}. The induced turbulence creates cloud-typical morphology and over-densities in the form of clumps and filaments. The high-density regions within these structures are potential sites of star formation \citep{2011A&A...529L...6A,2013A&A...551C...1S,2014prpl.conf...27A}. The velocity dispersion on the scale of turbulence driving is assigned as $\sigma_v=c_\mathrm{s}\,\mathcal{M} = 1.0\, \mathrm{km\, s^{-1}}$ such that the steady-state sonic Mach number $\mathcal{M}=5.0$.

The magnetic field is uniform initially with $B= 10^{-5}\, \mathrm{G}$ along the z-axis of the computational domain, but is later altered due to the tangling, stretching, and compression of magnetic field lines by the turbulence \citep{2021PhRvF...6j3701S}, producing a magnetic field structure similar to that observed in real MCs \citep{2016JPlPh..82f5301F}. The initial virial parameter is set as $\alpha_\mathrm{{vir}}=2E_\mathrm{{kin}}/E_\mathrm{{grav}}=0.5$ which is consistent with the observed values \citep{1992A&A...257..715F,2013ApJ...779..185K,2015ApJ...809..154H}. We analyse the statistical properties like the IMF and time evolution of different dynamical quantities of the formed stellar clusters from this point in time, which we set as $t = 0$, i.e., when self-gravity is turned on. Such a technique is analogous to that employed in previous studies \citep[e.g.,][]{2012ApJ...761..156F,2012ApJ...754...71K,2016ApJ...822...11P,2018MNRAS.480..182G,2020MNRAS.496.5201M,2021MNRAS.507.2448M}.

\section{Results}
\label{sec:results}

\begin{figure*}
    \centering
    \includegraphics[width=\textwidth]{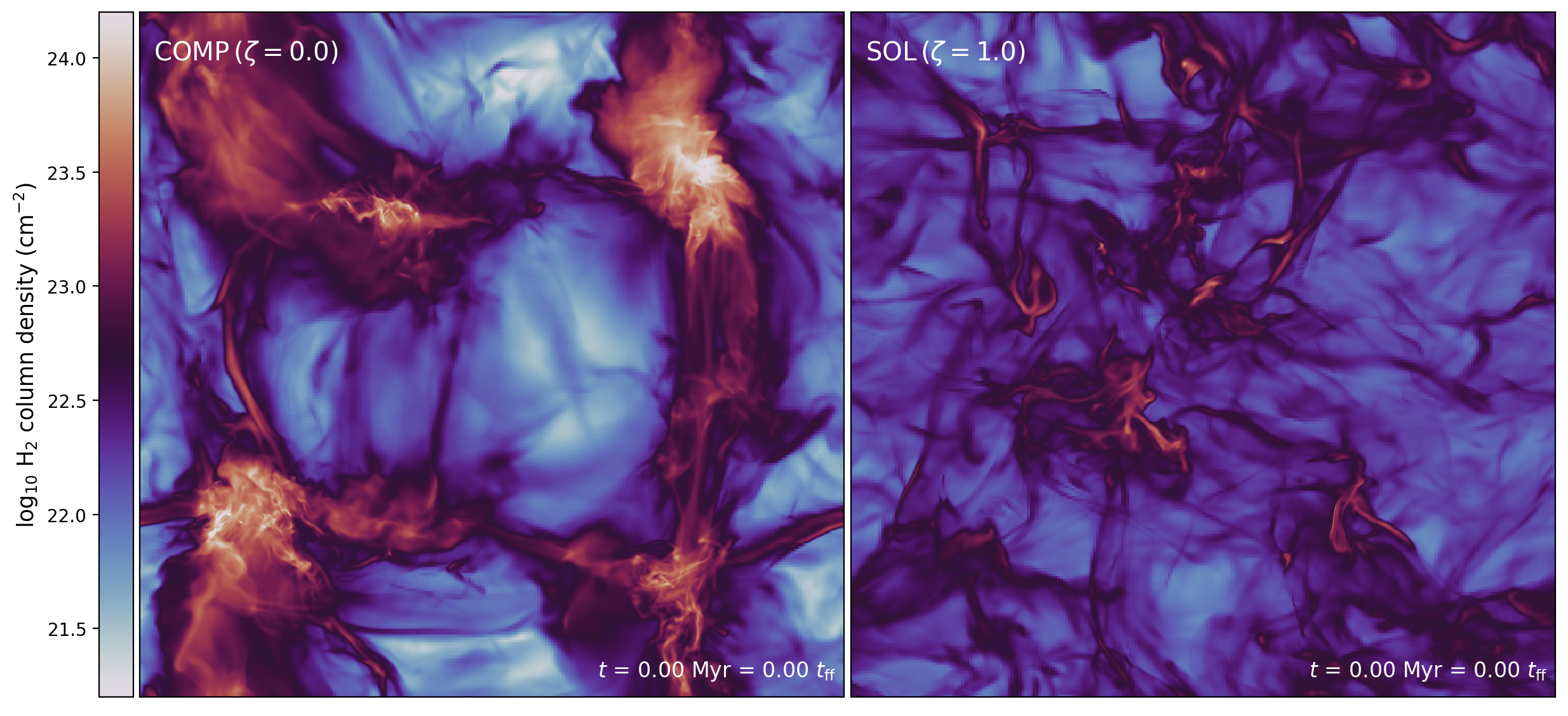}
    \caption{Left panel: The mass-weighted column density map of one of the simulations with a purely compressive driving (COMP) at the moment self-gravity is turned on, i.e., at $t=0$. Right panel: The mass-weighted column density map of a purely solenoidal driving (SOL) simulation with the same turbulence seed and at the same time.}
    \label{fig:denmap_comp_0005}
\end{figure*}

\begin{figure*}
    \centering
    \includegraphics[width=\textwidth]{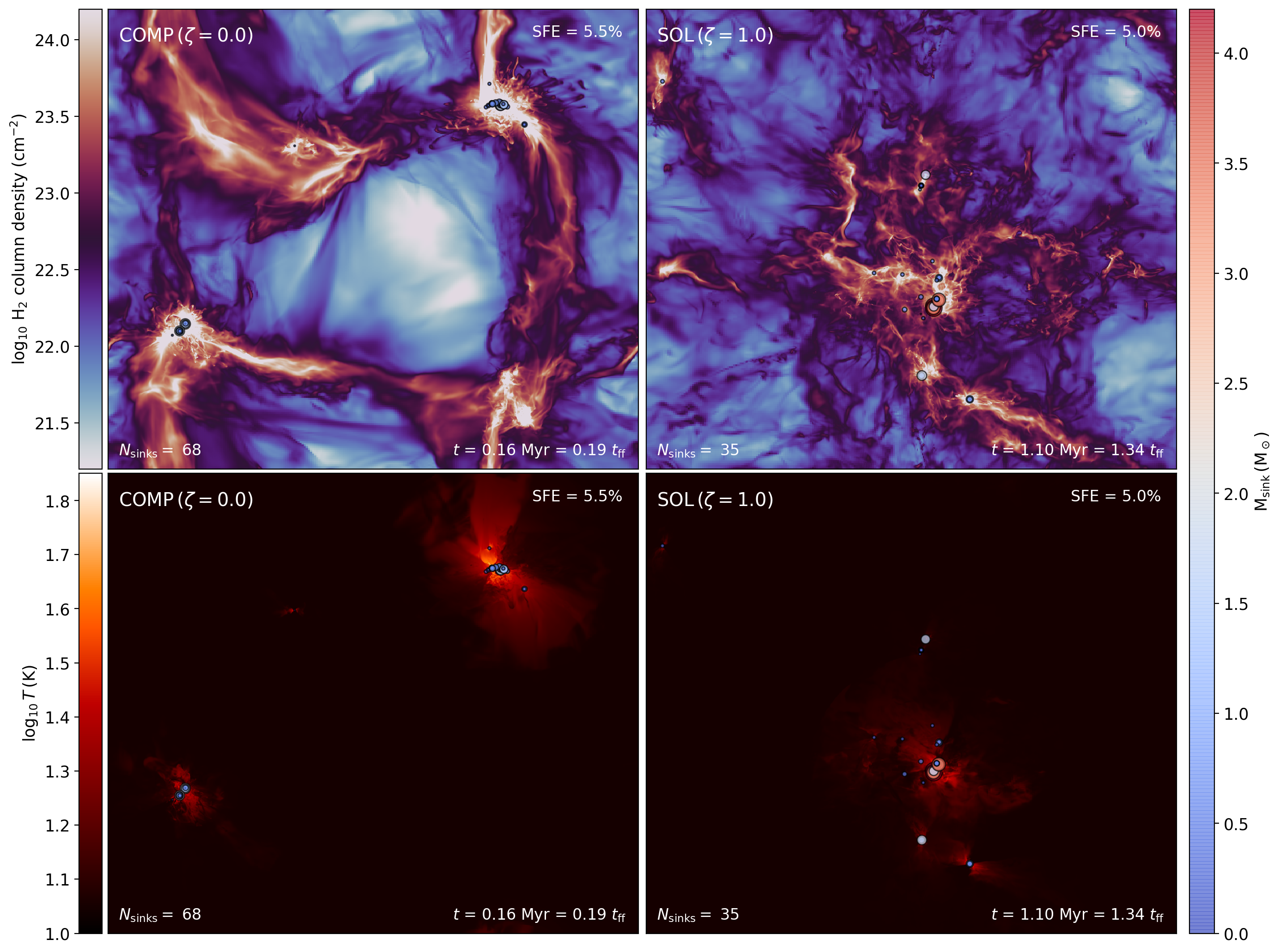}
    \caption{Left panel: Column density map (mass-weighted) of the COMP simulation shown in Fig.~\ref{fig:denmap_comp_0005} at a star formation efficiency (SFE) of $5\%$. Right panel: The mass-weighted column density map of the SOL simulation shown in Fig.~\ref{fig:denmap_comp_0005} at $\mathrm{SFE}=5\%$. The circular markers correspond to the sink particle (star+disc system) positions and the colour bar on the right represents the mass of the sink particles. The size of the markers is scaled by the mass of the sink particles.} 
    \label{fig:turbparam_mapcomp_sfe5}
\end{figure*}

We investigate the impact of the mode of turbulence driving by comparing MHD cloud-collapse simulations that use a purely compressive turbulence driving with simulations that are driven by purely solenoidal modes. To enhance the statistical significance, we carry out multiple simulations for each of the models with the same initial setup as prescribed in \S\ref{sec:parameters}, but with different realisations of the turbulent flow. For the purely compressive model (COMP), we perform 7~simulations with different turbulence realisations, and for the purely solenoidal model (SOL), we run a total of 11~simulations with different turbulence realisations, to ensure that the total number of sink particles (protostars) formed is comparable in both the COMP and SOL models. The left panel in Fig.~\ref{fig:denmap_comp_0005} shows the mass-weighted column density\footnote{We define the mass-weighted projection of the gas number density as $\int \rho^2\, dz\, / \int \rho\, dz$ and the mass-weighted projection of the temperature as $\int \rho T\, dz\, / \int \rho\, dz$, where the projection is taken along the $z$-direction. All figures in this paper depicting density and temperature maps are mass-weighted. The objective of the mass-weighting is to allow better visualisation of the morphological features, i.e., to highlight the densest structures.} of one of the COMP simulations and the right panel shows the same for the SOL simulation with the same turbulence realisation at the moment self-gravity is turned on. It can be clearly seen that the over-dense regions produced by turbulent shocks in the COMP model are comparatively larger in size and have a higher density on average than the over-dense structures in the SOL model. Therefore, as soon as self-gravity is turned on, star formation commences aggressively in COMP, while it is much slower in the SOL run. Fig.~\ref{fig:turbparam_mapcomp_sfe5} shows the mass-weighted column density (top row) and temperature structure (bottom row) of COMP (left column) and SOL (right column) simulations shown in Fig.~\ref{fig:denmap_comp_0005}, but at a star formation efficiency (SFE) of $5\%$. In the initial stages of the SOL simulation, a few stars form in some of the dense structures, but a substantial fraction of the stars form only much later when some of these structures merge under the action of self-gravity or due to the large-scale turbulent motions.

\begin{table*}
	\caption{Key simulation parameters and results.}
	\label{tab:sims}
	\begin{tabular}{lccccccc} % four columns, alignment for each
	    \hline
		\hline
		 Model & $N_{\mathrm{sims}}$ & $N_{\mathrm{Total\, sinks}}$ & $\overline{M}_{\mathrm{median}}\, [\mathrm{M_{\odot}}] $ & $\overline{M}_{\mathrm{avg}}\, [\mathrm{M_{\odot}}]$ & SSF \\
        (1) & (2) & (3) & (4) & (5) & (6)\\
		\hline
		\hline 
		COMP & $7$ & 468 & $0.4\pm0.1$ & $0.6\pm0.2$ &  $0.59\pm0.08$ \\
		SOL & $11$ & 445 & $0.6\pm0.2$ & $1.0\pm0.2$  & $0.64\pm0.09$ \\
		
		\hline
	\end{tabular}
	\\
    \raggedright\textbf{Notes.} Multiple simulations with different turbulence realisations are run for the compressive turbulence driving (COMP) and solenoidal turbulence driving (SOL) models. All values quoted in the table are calculated at SFE = 5\%. The resolution level and cloud properties are the same in both models, and the only difference is the mode of turbulence driving imposed. Main simulation parameters: computational box size: $L=2\, \mathrm{pc}$, uniform initial gas density: $\rho_{\circ} = 6.56 \times 10^{-21}\, \mathrm{g\, cm^{-3}}$, total cloud mass: $M_{\mathrm{cl}}=775\, \mathrm{M_{\odot}}$, uniform initial magnetic field: $B=10^{-5}\, \mathrm{G}$ (along the z-axis), velocity dispersion on the driving scale of the turbulence: $\sigma_v=1.0\, \mathrm{km\, s^{-1}}$, maximum effective grid resolution: $N_{\mathrm{eff,\, res}}^3=4096^3$ cells, minimum cell size: $\Delta x_{\mathrm{cell}}=100\,\mathrm{AU}$, and sink particle threshold density: $\rho_{\mathrm{sink}}=3.8\times10^{-16}\, \mathrm{g\ cm^{-3}}$.
\end{table*}

\subsection{Evolution of dynamical quantities}
\label{sec:evol_dynamicQ}
\begin{figure*}
    \centering
    \includegraphics[width=\textwidth]{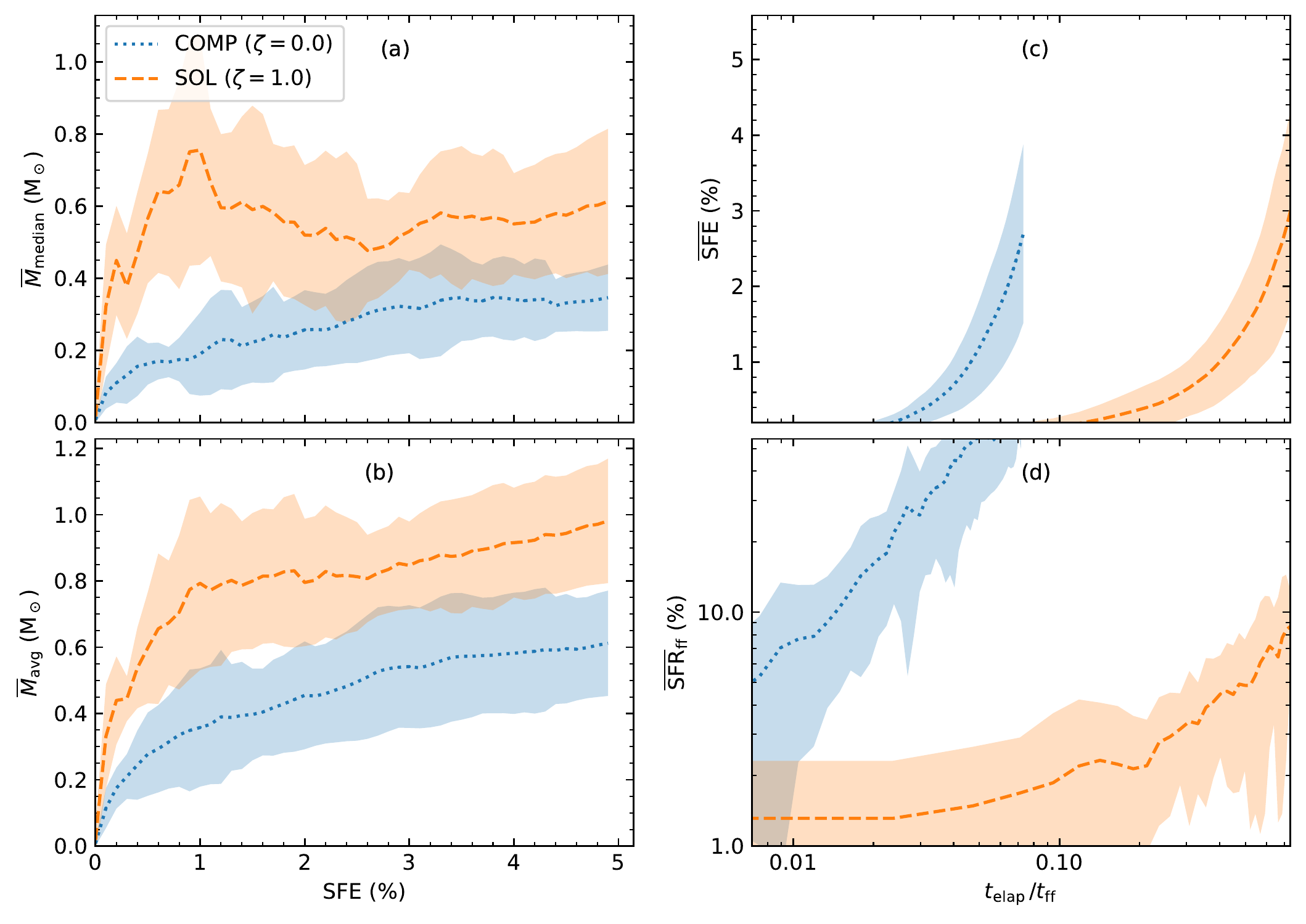}
    \caption{The left panels show (a) the median stellar mass and (b) the average stellar mass as a function of the star formation efficiency (SFE in \%) for the COMP (dotted curve) and SOL (dashed curve) simulations. The right panels (c) and (d) indicate the star formation efficiency and star formation rate per free-fall time, respectively, as a function of time. For both COMP and SOL models, all quantities shown here represent the average values obtained from multiple simulations, and the coloured bands correspond to the standard deviation over the set of these simulations. Here $t_{\mathrm{elap}}/t_{\mathrm{ff}}$ is the elapsed time from the formation of the first sink particle in units of the free-fall time and is distinguished from the time $t$ in the above column density projections, which is the time measured from the instant self-gravity was turned on.} \label{fig:dynamicQ_timelog}
\end{figure*}

Fig.~\ref{fig:dynamicQ_timelog} presents the evolution of the median sink mass $\overline{M}_{\mathrm{median}}$ (panel~a) and average sink mass $\overline{M}_{\mathrm{avg}}$ (panel~b) as a function of SFE (\%). The overbar in the plotted quantities denotes that the respective values are averaged over multiple simulations. We find that, for both the COMP and SOL models, $\overline{M}_{\mathrm{median}}$ and $\overline{M}_{\mathrm{avg}}$ are nearly constant beyond an SFE of $\sim 1.5\%$. It is evident that the $\overline{M}_{\mathrm{median}}$ and $\overline{M}_{\mathrm{avg}}$ of the SOL simulations are relatively higher. On taking the average over the SFE range 1.5--5\%, in the case of the COMP model, the median and average sink particle mass are $0.31\pm0.04\, \mathrm{M_\odot}$ and $0.53\pm0.06\, \mathrm{M_\odot}$, respectively, while they are $0.55\pm0.03\, \mathrm{M_\odot}$ and $0.87\pm0.05\, \mathrm{M_\odot}$, respectively, for the SOL model. The right panels in Fig.~\ref{fig:dynamicQ_timelog} depict the evolution of the star formation efficiency $\overline{\mathrm{SFE}}$ (panel~c) and star formation rate per free-fall time $\overline{\mathrm{SFR_{ff}}}$ (panel~d) with time. The star formation rate is around an order of magnitude higher in the COMP simulations, as seen in previous simulations \citep[e.g.,][]{2012ApJ...761..156F,2017MNRAS.465..105L}. The $\overline{\mathrm{SFR_{ff}}}$ in the SOL simulations is between 1--3\% for the most part of the cloud evolution, but increases towards the end. The acceleration in the $\overline{\mathrm{SFR_{ff}}}$ in the later stages is due to the increased efficiency of gravity in the cluster-forming regions in bringing the gas together and increasing the local density, allowing more stars to form \citep[see also][]{2011MNRAS.416.1436B,2015ApJ...808...48B,2018A&A...611A..88L,2021MNRAS.507.4335K}. The average star formation rate in the Milky Way is estimated to be $\sim 1$--$2\%$ per free-fall time \citep{2007ApJ...654..304K,2010ApJ...723.1019H,2012ApJ...745...69K,2013MNRAS.436.3167F,2013ApJ...778..133L,2015ApJ...806L..36S,2016ApJ...831...73V,2019FrASS...6....7K,2019MNRAS.488.1407K}, although the spread about the average value can be large \citep{2010ApJ...723.1019H,2016ApJ...833..229L,2016ApJ...831...73V,2016A&A...588A..29H,2017ApJ...841..109O}. \citet{2016ApJ...833..229L} measured the star formation rates per free-fall time in $191$ star-forming giant molecular cloud complexes in the Milky Way and find that the dispersion in the rates is $\sim 0.9\, \mathrm{dex}$ with values as low as $0.01\%$ to as high as $100\%$ per free-fall time \citep[see top left panel in Fig.~4 of][]{2016ApJ...833..229L}.\footnote{Note that star formation rates per free-fall time exceeding $100\%$ are possible, if a particular cloud region undergoes local compression due to dynamical effects, such as shocks, which leads to a star formation rate that exceeds the purely gravitational free-fall rate \citep{2012ApJ...761..156F}.} Therefore, both the high star formation rates seen in the COMP model and the low star formation rates seen in the SOL model are consistent with the star formation rates measured in Milky Way clouds, depending on the specific cloud or cloud region selected.

\subsection{Sink mass distribution}
\label{sec:smd}
\begin{figure}
    \centering
    \includegraphics[width=\columnwidth]{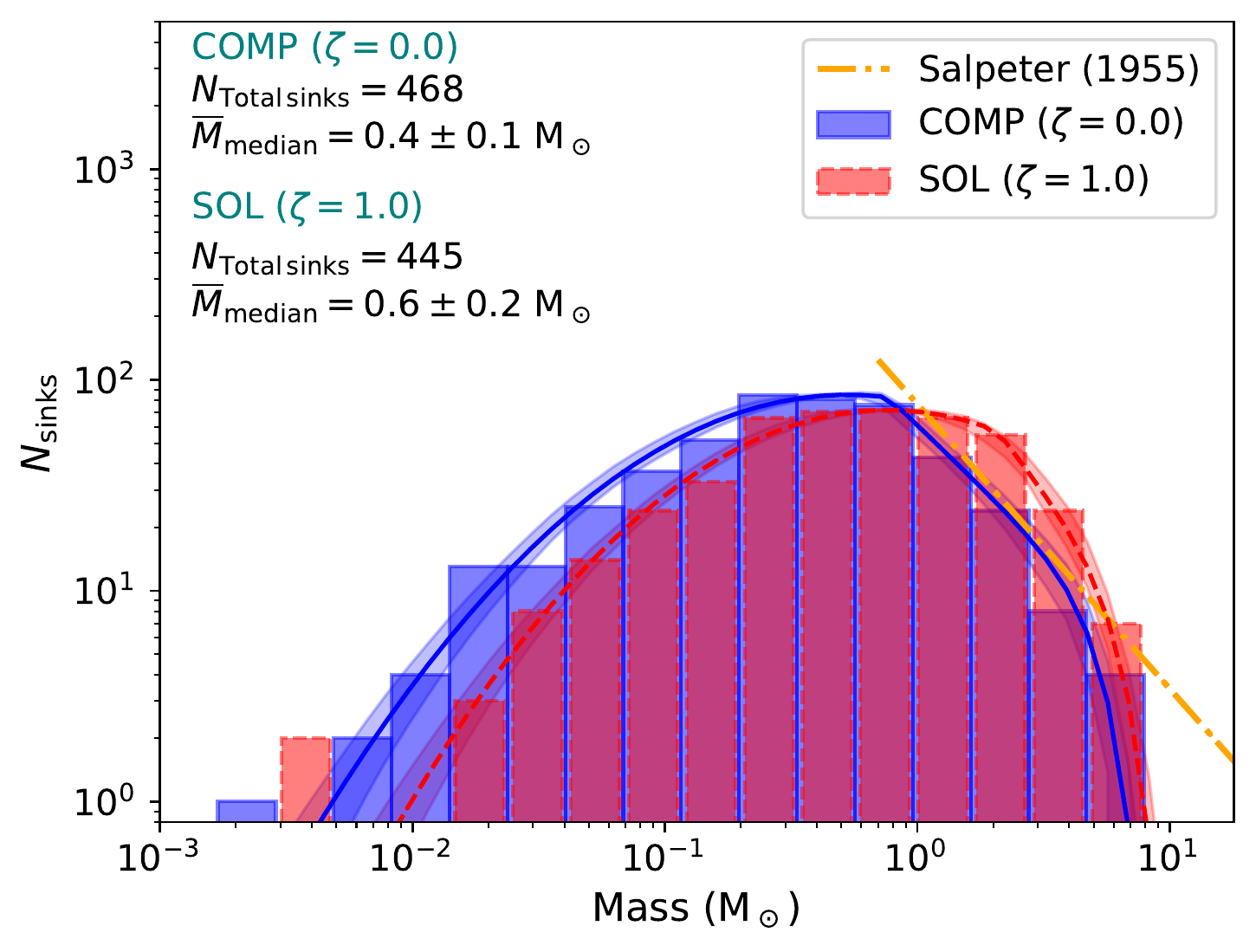}
    \caption{Comparison of the sink mass distribution (SMD) obtained for the COMP (histogram with solid edges) and SOL (histogram with dashed edges) turbulence driving models at SFE=5\%. The solid (COMP) and dashed (SOL) curves fitted (the 16th to 84th percentile confidence interval shown as the shaded region), are \citet{2005ASSL..327...41C}-type IMFs, but modified to take into account the finite mass of the simulated cloud (high-mass cutoff). The values of the IMF fit parameters (peak, standard deviation, transition mass and power-law slope) are derived using MCMC sampling (see \S\ref{sec:smd}). The dash-dotted line is the \citet{1955ApJ...121..161S} IMF.
    }  
    \label{fig:IMF_both}
\end{figure}

Fig.~\ref{fig:IMF_both} is a comparison between the sink mass distributions (SMDs) obtained for the COMP and SOL models at SFE = 5\%. The mass distributions represent data collected from multiple simulations with different turbulent realisations. We see that a change in the mode of turbulence driving affects the IMF considerably. The SOL SMD has a higher fraction of high-mass stars ($M_{\mathrm{sink}}> 1\, \mathrm{M_\odot}$) and has a slightly higher turnover (peak) mass. The median stellar mass of our COMP SMD is $ 0.4\pm0.1\, \mathrm{M_\odot}$ (at SFE = 5\%), while the same for the SOL SMD is $ 0.6\pm0.2\, \mathrm{M_\odot}$ (see Tab.~\ref{tab:sims}). We performed a KS test and obtained a p-value of the order of $10^{-8}$, meaning that we can neglect the hypothesis that the two distributions are identical.

\begin{table*}
\caption{Parameter values from the MCMC fit.}
\renewcommand{\arraystretch}{1.5}
\label{tab:mcmc}
\begin{tabular}{cccccccc}
\hline
\hline
& Model & $M_0\, [\mathrm{M_\odot}]$ & $\sigma$ & $M_{\mathrm{T}}\, [\mathrm{M_\odot}]$ & $\Gamma$ & $M_{\mathrm{cut}}\, [\mathrm{M_\odot}]$ & $p$ \\
& (1) & (2) & (3) & (4) & (5) & (6) & (7) \\
\hline
\hline
\multirow{2}{*}{Free $M_{\mathrm{T}}$} & COMP & $0.53_{-0.12}^{+0.12}$ & $0.68_{-0.05}^{+0.05}$ & $0.77_{-0.11}^{+0.19}$ & $1.2_{-0.2}^{+0.2}$ & $5.7_{-0.7}^{+0.9}$ & 4 \\
& SOL & $0.76_{-0.12}^{+0.21}$ & $0.64_{-0.05}^{+0.06}$ & $2.07_{-0.54}^{+0.85}$ & $1.5_{-0.5}^{+0.7}$ & $6.7_{-0.7}^{+0.9}$ & 4 \\
\hline
\multirow{2}{*}{Fixed $M_{\mathrm{T}}$} & COMP & $0.47_{-0.07}^{+0.12}$ & $0.67_{-0.05}^{+0.06}$ & 1 & $1.4_{-0.2}^{+0.2}$ & $6.2_{-0.9}^{+1.1}$ & 4 \\
& SOL & $0.85_{-0.13}^{+0.10}$ & $0.65_{-0.04}^{+0.04}$ & 1 & $0.8_{-0.2}^{+0.1}$ & $6.2_{-0.6}^{+0.8}$ & 4 \\
\hline
\end{tabular}
\\
\raggedright\textbf{Notes.} The values presented here correspond to the $50^{\mathrm{th}}$ percentile of each of the parameters, with the $16^{\mathrm{th}}$ and $84^{\mathrm{th}}$ percentiles denoting the uncertainty.
\end{table*}

To quantitatively confirm that the apparent difference between the shape of the COMP and SOL SMDs is not a consequence of the binning choice, we fit a modified version of the \citet{2005ASSL..327...41C} IMF to our SMD data using the Markov Chain Monte-Carlo (MCMC) sampler emcee of \citet{2013PASP..125..306F} \citep[see also][]{2021MNRAS.503.1138N}. As opposed to other common model fitting methods, the MCMC sampling technique does not require binning of data. In order to account for the finite mass in our computational box, we include an exponential term that acts as a smooth cutoff at high masses in the power law part of the \citet{2005ASSL..327...41C} IMF,
 \begin{equation}
    dN / d\mathrm{log} M =
      \begin{cases}
        k_1\,\mathrm{exp}\,\left(-\frac{(\mathrm{log}\, M - \mathrm{log}\, M_0)^2}{2\, \sigma^2}\right)   & \text{for\, $M < M_{\mathrm{T}}$,}\\ 
        k_2\,M^{-\Gamma}\, \mathrm{exp}\,\left(-(M/M_{\mathrm{cut}})^{p}\right)  & \text{for\, $M \geq M_{\mathrm{T}}$.}\\
      \end{cases}
      \label{Eq: Chab}
\end{equation}
with five free parameters $\theta = (\mathrm{log}\, M_0, \sigma, \mathrm{log}\, M_{\mathrm{T}}, \Gamma, \mathrm{log}\, M_{\mathrm{cut}})$ where $M_0, \sigma, M_{\mathrm{T}}$ and $\Gamma$ are the peak mass, standard deviation of the log-normal part, mass at which the IMF transitions from a log-normal to a power-law form, and slope of the power-law part, respectively. $k_1$ and $k_2$ are normalisation constants, set to ensure continuity at $M_{\mathrm{T}}$. Due to the presence of the exponential term in the power-law part, the IMF will be cut-off at high masses. The mass at which the exponential term starts to dominate is characterised by $M_{\mathrm{cut}}$ and $p$ defines how sharply the IMF drops around $M_{\mathrm{cut}}$.  

The posterior probability $P(\theta | \{M_{\mathrm{sink}}\})$, i.e., the probability of $\theta$ given the list of sink particle masses \{$M_{\mathrm{sink}}$\} can be calculated using the Bayes' theorem and is given by
\begin{equation}
    P(\theta | \{M_{\mathrm{sink}}\}) = \frac{P(\theta)\, P(\{M_{\mathrm{sink}}\} | \theta)}{\int{P(\theta')\, P(\{M_{\mathrm{sink}}\} | \theta')\, d\theta'}},
\end{equation}
where $P(\theta)$ represents the prior distribution and $P(\{M_{\mathrm{sink}}\} | \theta)$ is the likelihood function, i.e., probability of \{$M_{\mathrm{sink}}$\} given the IMF form defined by Eq.~(\ref{Eq: Chab}) with a particular parameter combination $\theta$. The likelihood function is given by \citep{2021MNRAS.503.1138N} 
\begin{equation}
    P(\mathrm{\{M_{\mathrm{sink}}\}} | \theta) = \prod_{M_{\mathrm{i}} \epsilon \{M_{\mathrm{sink}}\}}\, \frac{dN}{dM}(M_{\mathrm{i}},\theta).
\end{equation}

We employ uniform priors on $\mathrm{log}\, M_0, \sigma, \mathrm{log}\, M_{\mathrm{T}}$ and $\Gamma$. We set $p = 4$ since we want the cut-off to be sufficiently sharp. We note that changing $p$ in the range from $1$ to $10$ does not affect the fit of the relevant physical quantities, most importantly, $\mathrm{log}\,M_0, \sigma, \mathrm{log}\,M_{\mathrm{T}}$, and $\Gamma$. In the case of the parameter $\mathrm{log}\, M_{\mathrm{cut}}$, we need to be cautious while defining the prior. Due to the low statistics in the high-mass end of our SMDs, the error in estimating $\mathrm{log}\, M_{\mathrm{cut}}$ can be large. Therefore, we need to have a rough estimate of where $\mathrm{log}\, M_{\mathrm{cut}}$ is located. Accordingly, instead of a uniform prior, we use a Gaussian prior on $\mathrm{log}\, M_{\mathrm{cut}}$ with the mean of the Gaussian defined by the maximum sink particle mass $M_{\mathrm{max}}$ in our simulations. For deriving $M_{\mathrm{max}}$, first the MCMC fit is derived as discussed above, except with a uniform prior on $\mathrm{log}\, M_{\mathrm{cut}}$. The fit thus obtained for each of the driving models correspond to the mass distribution of sink particles obtained from multiple simulations. The fit is then rescaled to correspond to a single simulation by dividing by the total number of simulations. $M_{\mathrm{max}}$ will be the mass at which the number of stars is less than 1 in the rescaled fit obtained with uniform priors. Finally, the MCMC fitting is performed again using a Gaussian prior for $\mathrm{log}\,M_{\mathrm{cut}}$ with a mean of $\mathrm{log}\,M_{\mathrm{max}}$. Using this method, we find a stable value for $\mathrm{log}\,M_{\mathrm{cut}}$ automatically, without having to impose any prior knowledge of its final value. Most importantly, while the cutoff allows us to account for the fact that our simulated clouds have a finite mass, $\mathrm{log}\,M_{\mathrm{cut}}$ is sufficiently high that none of the main physical parameters are affected by its details, namely $\mathrm{log}\,M_0, \sigma, \mathrm{log}\,M_{\mathrm{T}}$ and $\Gamma$.

The corner plot showing the posterior probability distribution of the parameters is presented in the Appendix section~\ref{sec:appendix}. Tab.~\ref{tab:mcmc} lists the $50^{\mathrm{th}}$ percentile value of the parameters obtained using the MCMC technique. The error bars denote the $16^{\mathrm{th}}$ and $84^{\mathrm{th}}$ percentiles. The parameter set obtained for the COMP and SOL SMDs are clearly different. The COMP model has a lower $M_0$ and also a slightly higher $\sigma$, which indicates the presence of a higher fraction of low-mass stars as compared to the SOL model. In addition, the combination of $M_0$ and $M_{\mathrm{T}}$, which controls how the IMF turns over from a log-normal form to a power law, varies between the two models. While $M_0=0.5\, \mathrm{M_\odot}$ and $M_{\mathrm{T}}=0.8\, \mathrm{M_\odot}$ in the case of the COMP model, they are located at $0.8\, \mathrm{M_\odot}$ and $2.1\, \mathrm{M_\odot}$, respectively, for the SOL model. The solid curve in Fig.~\ref{fig:IMF_both} corresponds to the fit derived for the COMP SMD using the $50^{\mathrm{th}}$ percentiles of each of the parameters, with the spread bracketed by the $16^{\mathrm{th}}$ and $84^{\mathrm{th}}$ percentiles. The dashed curve represents the same for the SOL SMD. The curves compare very well with the corresponding histograms, justifying our binning choice and confirming that the SMDs produced with the two driving modes are different. 

Our simulations do not produce very high-mass stars and the cut-off mass $M_{\mathrm{cut}}$ occurs well before $10\, M_\odot$ in both the models (see Tab.~\ref{tab:mcmc}). As a consequence of the narrow high-mass range, it is difficult to have an accurate estimate of the power-law slope, which is why the error bars on $\Gamma$ are large, particularly for the SOL model. In such a situation, small variations in the location of the transition mass $M_{\mathrm{T}}$ can significantly affect the value of the power-law slope. To understand the uncertainties that this introduces, we also produce another set of fits for our SMDs using MCMC sampling in the same manner as discussed above, but with $M_{\mathrm{T}}$ fixed at the transition mass for a \citet{2005ASSL..327...41C} IMF, i.e., at $1\, \mathrm{M_\odot}$. The corresponding parameter values are shown in Tab.~\ref{tab:mcmc}. We see that on fixing $M_{\mathrm{T}} = 1\, \mathrm{M_\odot}$, there is no significant change in the parameter values that define the log-normal part of the IMF fit, namely, $M_0$ and $\sigma$. However, we find that for the COMP fit, $\Gamma$ becomes slightly steeper compared to its value when $M_{\mathrm{T}}$ was a free parameter (although not statistically significant, i.e., a change from $\Gamma=1.2$ to $1.4$, which is within the 1-sigma uncertainty), but for the SOL case, $\Gamma$ becomes significantly shallower (from $1.5$ to $0.8$, just outside a 1-sigma overlap, considering the uncertainties of both fits). When $M_{\mathrm{T}}$ was a free parameter, the value of $M_{\mathrm{T}}$ derived for the COMP fit was lower than $1\, \mathrm{M_\odot}$, while it was higher than $1\, \mathrm{M_\odot}$ for the SOL fit. Therefore, on fixing $M_{\mathrm{T}}$ at $1\, \mathrm{M_\odot}$, $M_{\mathrm{T}}$ moves further away from the peak $M_0$ in the case of the COMP fit, while it moves closer to $M_0$ in the case of the SOL fit. This explains why $\Gamma$ becomes steeper for the COMP fit and shallower in the case of the SOL fit. The combination of $M_{\mathrm{T}}$ and $\Gamma$ obtained when $M_{\mathrm{T}}$ is a free parameter and those obtained when $M_{\mathrm{T}}$ is fixed both qualitatively agree on the fact that the SOL SMD has a higher fraction of high-mass stars. The IMF fits obtained with the parameter values for the fixed $M_{\mathrm{T}}$ case shown in Tab.~\ref{tab:mcmc} (see Fig.~\ref{fig:imf_both_mt_fixed}) and the associated parameter correlation (corner) plots (see Fig.~\ref{fig:cornerplot_blue_mt_fixed} and Fig.~\ref{fig:cornerplot_red_mt_fixed}) are presented in the Appendix section~\ref{sec:appendix}.

\begin{figure*}
    \centering
    \includegraphics[width=\textwidth]{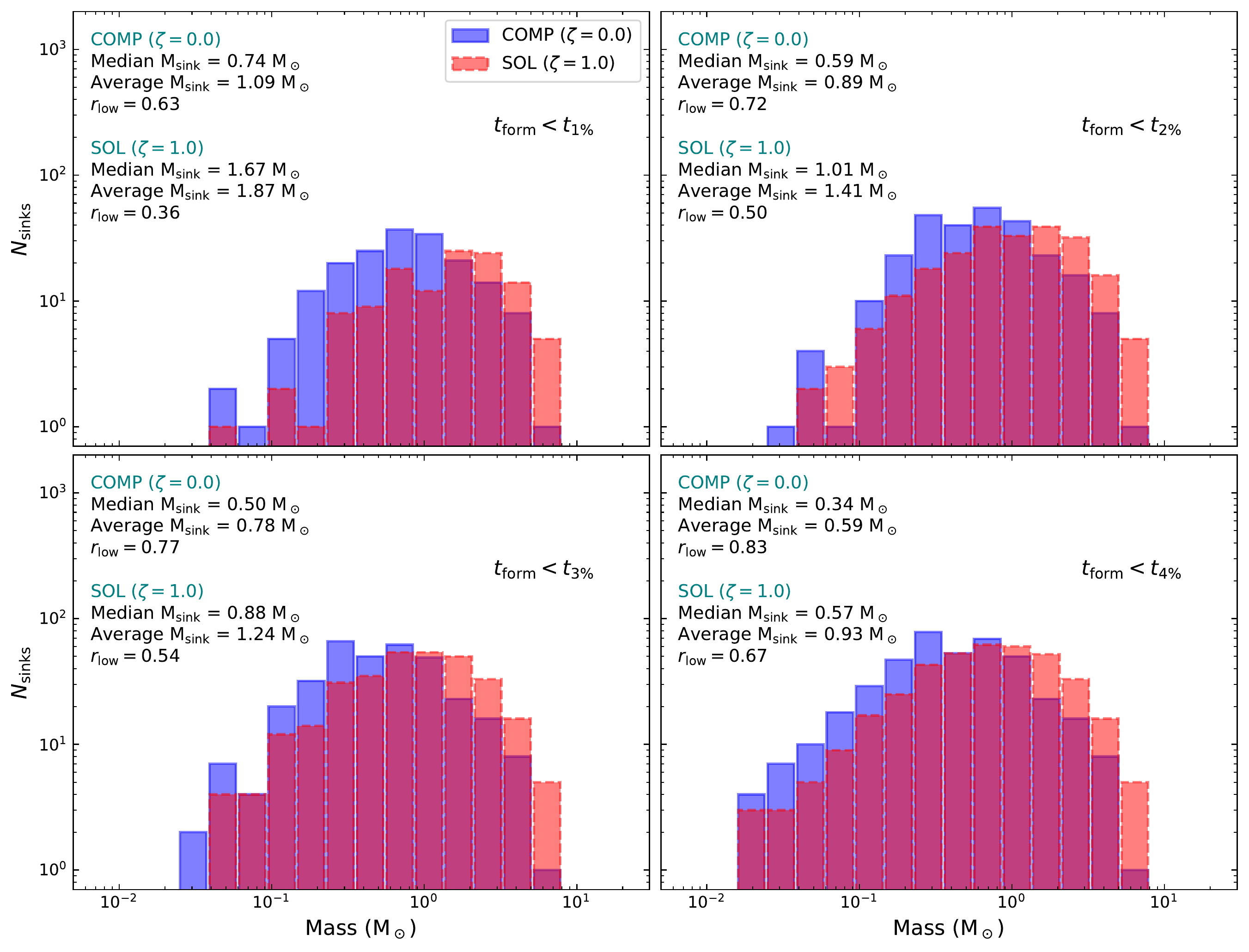}
    \caption{Distribution of stellar masses (mass at the end of the simulation, i.e., at SFE=5\%) of all the sink particles that formed before an SFE of (from top left to bottom right) 1\%, 2\%, 3\% and 4\% is reached. The histogram with solid edges represents the COMP distribution and the histogram with dashed edges corresponds to the SOL distribution. We point out that for calculating the median and average sink masses shown in the plots here, the sink particle masses at SFE = 5\% are used, but only of the sink particles that formed before a particular SFE. On the other hand, $\overline{M}_{\mathrm{median}}$ in Fig.~\ref{fig:IMF_both} represents the median value for the sink particle masses at SFE = 5\%, where all sink particles are considered, irrespective of when they formed.}
    \label{fig:imf_evol}
\end{figure*}

The plots shown in Fig.~\ref{fig:imf_evol} present the sink mass distribution at SFE = 5\%, but only of the sink particles that formed before the time at which a particular SFE is reached. For example, the top left panel shows the mass distribution of sink particles that formed before an SFE of 1\% is reached, while the bottom right panel shows the mass distribution of the sinks that formed before an SFE of 4\% is reached. We note that Fig.~\ref{fig:imf_evol} does not represent the time evolution of the SMD, i.e., the distribution of stellar masses at different SFEs, rather it shows the distribution of final stellar masses (mass at the simulation end time, i.e., at SFE = 5\%) of all the sink particles that were created before an SFE of 1\%, 2\%, 3\% and 4\% (from top left to bottom right panel in Fig.~\ref{fig:imf_evol}) is reached. We see that the peak of the distribution shifts to lower masses as we progressively include stars that form at later times. This is readily seen for SOL, where the peak is at around $2-3\,\mathrm{M_\odot}$ when only sinks that form before SFE=1\% are included, while it is  $\sim 0.5-1.0\,\mathrm{M_\odot}$ in the mass distribution when all the sink particles are included, i.e., sink particles that form before SFE = 5\% (see Fig.~\ref{fig:IMF_both}). There is also a shift in the peak of the COMP model, although relatively minor, from $\sim 0.7-0.9\, \mathrm{M_\odot}$ to $\sim 0.3-0.5\, \mathrm{M_\odot}$. The shift to lower masses implies that the formation of comparatively lower-mass stars is more favourable at later times, which is also indicated by the decrease in the median and average mass as we include more younger stars in the distribution.

It is clear that the COMP and SOL distributions differ the most in the top left panel of Fig.~\ref{fig:imf_evol}, where only the sink particles that form in the early stages are considered. This is because these sink particles form before self-gravity modifies the gas density PDF substantially and begins to dominate in promoting fragmentation. Therefore, the effect of the turbulence driving in setting the mass would be more pronounced for stars that form relatively early in the evolution of the cloud. For SOL, $r_{\mathrm{low}}$, the fraction of low-mass stars ($\mathrm{M < 1\, M_\odot}$) that form early in the simulation is substantially smaller compared to that in COMP, and the shape of the SOL SMD is not established when only the sink particles that form before SFE=1\% are included. The low-mass part of the SOL IMF becomes fully developed only when we include the younger sink particles, i.e., the sink particles that form in the later stages of the cloud evolution. The formation of many low-mass stars and the increase in the $\overline{\mathrm{SFR_{ff}}}$ (see panel~d in Fig.~\ref{fig:dynamicQ_timelog}) towards the end of the simulations are inter-related. At later stages, the average density of the gas in the star-forming regions (where a cluster of stars forms) increases as a result of the increased influence of self-gravity on fragmentation. Previous studies have shown that, mathematically, this corresponds to the development of a power-law tail in the gas density PDF \citep[e.g.,][]{2011ApJ...727L..20K,2011MNRAS.416.1436B,2013ApJ...763...51F,2015ApJ...808...48B,2018A&A...611A..88L,2021MNRAS.507.4335K}. Therefore, gravitationally induced fragmentation \citep{2018A&A...611A..88L} begins to play a more important role in setting the mass of the sink particles that form during the later stages. Further, as more stars form, the stellar density increases. As a result, the frequency of dynamical encounters rises and thus the sink particles that form in the later stages are prone to the termination of accretion early on via dynamical ejections \citep{2001AJ....122..432R,2002MNRAS.332L..65B}. The fragmentation induced by self-gravity and dynamical effects allow more low-mass stars to form.
 
\subsubsection{IMF in the Galactic centre}
The above discussion implies that clouds that are primarily driven by solenoidal modes will produce only a small fraction of low-mass stars (i.e., low fragmentation) if star formation in later stages is suppressed. Low fragmentation would automatically lead to the existing stars reaching high masses. Such a scenario is a possibility in the case of star-forming regions near the Galactic centre. The clusters within the Central Molecular Zone (CMZ), particularly near the Galactic centre, are found to have top-heavy IMFs, i.e., a higher fraction of high-mass stars compared to the typical IMF \citep{1999ApJ...525..750F,2006ApJ...653L.113K,2013ApJ...764..155L,2019ApJ...870...44H}. The turbulence driving in the CMZ is expected to be dominated by solenoidal modes as a result of the enhanced shear \citep{2016ApJ...832..143F,2022arXiv220613442R} in the CMZ environment. Thus, if the turbulence driving is primarily solenoidal, the deviation from the average density is small (relatively narrow gas density PDF), and therefore the formation of stellar masses lower than the mean Jeans mass is also expected to be small. This means that only few low-mass stars can form in the early stages of star formation in CMZ clouds. Since the average temperature in the CMZ is significantly higher than that in typical clouds located in the Galactic disc \citep{2016A&A...586A..50G}, the mean Jeans mass will also be high, which again suppresses the formation of low-mass stars \citep{2006MNRAS.368.1296B,2007MNRAS.374L..29K}. As shown above, low mass stars can form only later in solenoidally-driven star-forming regions when the local density increases as a result of the increase in the gravitational influence. However, by that time, the existing stars will have already grown to high masses, because of the high Jeans mass. The radiative heating by these highly luminous stars prevents further fragmentation, and thus the fraction of high-mass stars in CMZ clouds would be relatively higher than that in typical Milky way clouds. Thus, the predominately solenoidal turbulence driving in CMZ clouds may (at least in part) explain observations of a top-heavy IMF in the CMZ.

% =======================================
\section{Comparison of the SMDs with observational data and theoretical models}
\label{sec:IMF_comp}
\subsection{Comparisons with observed IMFs}

\begin{figure}
    \centering
    \includegraphics[width=\columnwidth]{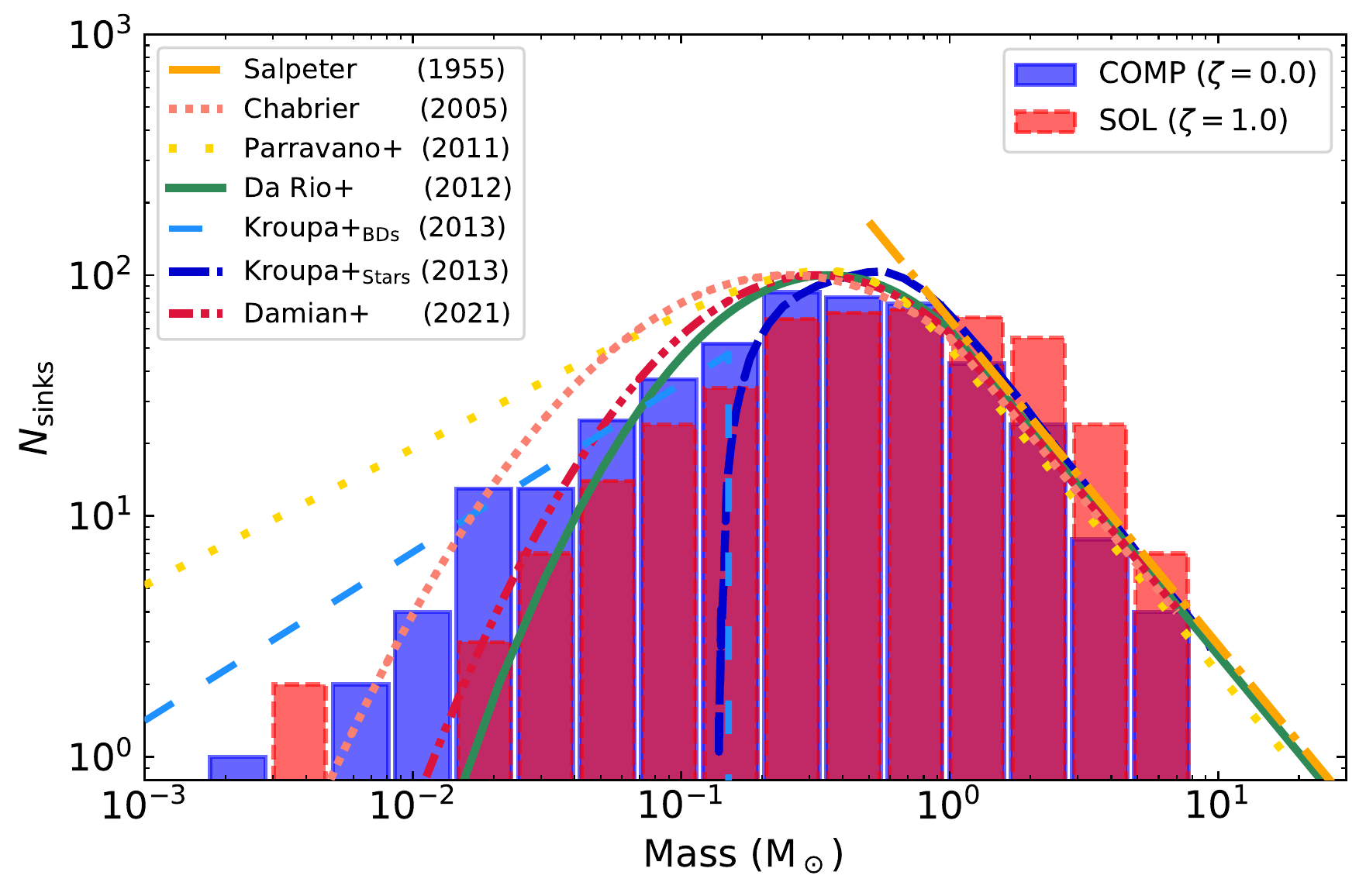}
    \caption{Comparison of various observational IMFs with the SMD (at SFE = 5\%) obtained in the COMP (histogram with solid edges) and SOL (histogram with dashed edges) simulation models. The curves are the system IMF models, based on observational surveys, by \citet{1955ApJ...121..161S} (dash-dotted), \citet{2005ASSL..327...41C} (short-dotted), \citet{2011ApJ...726...27P} (long-dotted), \citet{2012ApJ...748...14D} (solid), \citet{2013pss5.book..115K} for brown dwarfs (long-dashed) and stars (short-dashed), and \citet{2021MNRAS.504.2557D} (dash-dot-dotted).}  
    \label{fig:imf_obs}
\end{figure}

In Fig.~\ref{fig:imf_obs}, we compare the SMD of each driving model with IMF fits obtained in different observational surveys since \citet{1955ApJ...121..161S} (dash-dotted line). We compare the SMDs with the system IMFs instead of the canonical or the individual-star IMFs, because fragmentation on very small scales is not well resolved in our simulations, and therefore we cannot identify all the low-order multiple systems. The short-dotted curve in Fig.~\ref{fig:imf_obs} represents the \citet{2005ASSL..327...41C} system IMF. \citet{2011ApJ...726...27P} propose an analytical model of the IMF (long-dotted line) described by various parameters based on observational constraints, e.g., the ratio of the number of brown dwarfs (BDs) to the number of stars and the slope of the high-mass regime of the IMF \citep[see also][]{2000ApJ...534..870P}. This function predicts a higher fraction of BDs below $0.03\, \mathrm{M_\odot}$ than the \citet{2005ASSL..327...41C} IMF. \citet{2012ApJ...748...14D} (solid line) fit a log-normal function for the mass distribution of low-mass stars in the Orion Nebula Cluster (ONC). However, the standard deviation of their fit is smaller compared to that in \citet{2005ASSL..327...41C}, i.e, they find a lower fraction of brown dwarfs as compared to that found in the Galactic disc. We adopt the best-fit parameters, namely the characteristic mass $m_c$ and the standard deviation $\sigma\, (\mathrm{log\,} m)$, from table~3 of \citet{2012ApJ...748...14D} to reproduce their log-normal fit to the mass distribution they derived by considering a \citet{1998A&A...337..403B} evolutionary model. We then extend the fit to higher masses by combining it with a Salpeter-like power-law function, similar to what was done in \citet{2012ApJ...754...71K}. The \citet{2013pss5.book..115K} system IMF has separate mass functions for stars and BDs based on the argument that if BDs form in the same manner as stars, then it contradicts the observed binary properties of BDs. The \citet{2013pss5.book..115K} stellar system IMF \citep[taken from Fig.~25 in][]{2013pss5.book..115K} and BD IMF (short-dashed and long-dashed lines) are obtained by random pairing of companions out of the canonical IMF \citep{2001MNRAS.322..231K}, where initial binary fractions of 100\% and 0\%, respectively, are assumed. \citet{2021MNRAS.504.2557D} (dash-dot-dotted line) compared the stellar mass distribution of nine young clusters with different environmental conditions with respect to the number of massive stars, stellar density and the Galacto-centric distance. They found that the functional form of the distributions are relatively similar and that they can be fitted by a log-normal distribution with a peak at $0.32\pm0.02$ and $\sigma=0.47\pm0.02$ (in logarithmic scale).

The peak of the COMP SMD is at around $0.3-0.5\, \mathrm{M_\odot}$, while the peak of the SOL SMD lies between $0.5-1.0\, \mathrm{M_\odot}$. The \citet{2005ASSL..327...41C} (dotted line) and \citet{2012ApJ...748...14D} (solid line) system IMFs have peak masses of $\sim 0.25\, \mathrm{M_\odot}$ and $0.35\, \mathrm{M_\odot}$, respectively. The peak of the COMP SMD is comparable with the peak of the IMF derived from different observational surveys, which is located at around $\sim0.3\, \mathrm{M_\odot}$ \citep{2010ARA&A..48..339B,2014prpl.conf...53O}. However, the peak of the SOL SMD is too high even when considering the scatter in the observational estimates. Observational surveys (where close binaries are unresolved) find that approximately one BD is formed per every five late-type (sub-solar) stars \citep{2006AJ....132.2296A,2008ApJ...683L.183A,2007ApJ...671..767T,2011ApJ...726...27P,2013pss5.book..115K}. The ratio of the number of sink particles with sub-stellar masses ($M_{\mathrm{sink}} \leq 0.08\, \mathrm{M_\odot}$) to that of the sink particles with stellar masses ($0.15\,\mathrm{M_\odot} < M_{\mathrm{sink}} \le 1.0\,\mathrm{M_\odot}$) are $67/278 = 0.24$ and $30/235 = 0.13$ for the COMP and SOL models, respectively. Our results imply that variations in the IMF, e.g., the discrepancy in the width of the low-mass end between the different observational IMF models, may be explained by different mixtures of turbulence driving modes in the ISM.

\subsection{Comparisons with theoretical models of the IMF}

\begin{figure}
    \centering
    \includegraphics[width=\columnwidth]{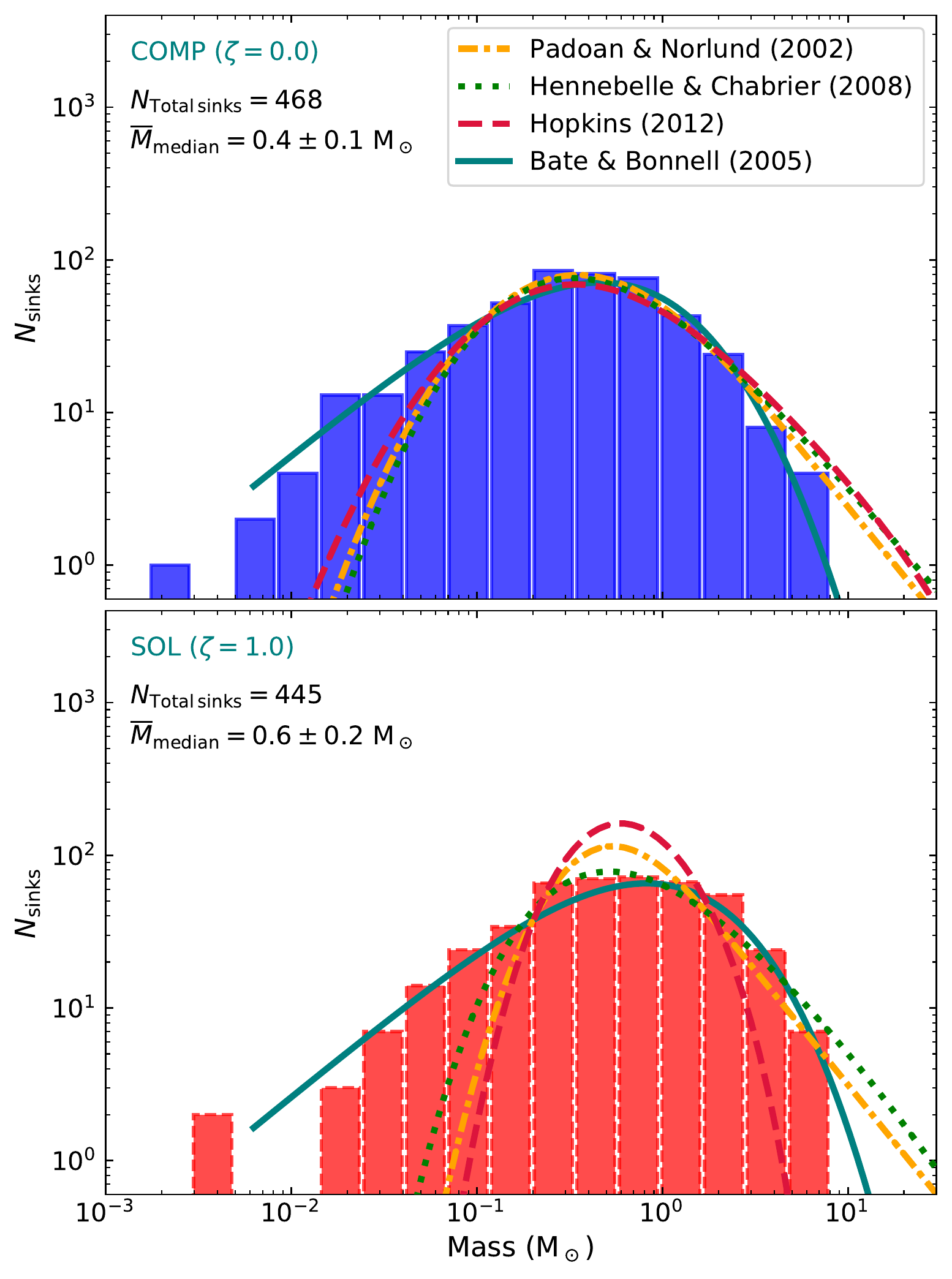}
    \caption{Top panel: Comparison between the sink mass distribution of the COMP model at SFE = 5\% with different theoretical models of the CMF/IMFs. The plotted curves correspond to \citet{2002ApJ...576..870P} (dash-dotted), \citet{2005MNRAS.356.1201B} (solid), \citet{2008ApJ...684..395H} (dotted), and \citet{2012MNRAS.423.2037H} (dashed) CMF/IMFs. Both the \citet{2002ApJ...576..870P} and \citet{2012MNRAS.423.2037H} CMF/IMFs have been shifted to lower masses by a factor of 1.3 and the \citet{2008ApJ...684..395H} CMF/IMF has been shifted to higher masses by a factor of 2, so as to fit the SMD. Bottom panel: Same as the top panel, but for the SOL simulations. Here the \citet{2002ApJ...576..870P}, \citet{2008ApJ...684..395H}, and \citet{2012MNRAS.423.2037H} CMF/IMFs have been shifted to lower masses by a factor of 2.5, 1.7 and 8.7, respectively.} 
    \label{fig:imf_theo}
\end{figure}

\subsubsection{The \citet{2002ApJ...576..870P} (PN02) model}
\citet{2002ApJ...576..870P} proposed that the size of cores that form in shocked regions of clouds created by supersonic turbulence is comparable to the thickness of the shocked layers. Assuming the isothermal shock jump conditions, the mass of a dense core is found to be inversely proportional to the square of the Mach number, which in turn is scale-dependent, following the Larson relation \citep{1981MNRAS.194..809L}. Taking into consideration the scale dependence of the Mach number, or equivalently, the power-law nature of the velocity power spectrum $P(k)\propto k^{-\beta}$ \citep{2021NatAs...5..365F}, \citet{2002ApJ...576..870P} arrive at the mass distribution of dense cores given by
\begin{equation}
    N(M)\, d\mathrm{log}M \propto M^{-3/(4-\beta)} d\mathrm{log}M.
\end{equation}
They argue that the distribution of collapsing cores is given by
\begin{equation}
    N(M)\, d\mathrm{log}M \propto M^{-3/(4-\beta)} \left(\int_{0}^{M} p(M_{\mathrm{J}})\, \mathrm{d}M_{\mathrm{J}}\right)\, d\mathrm{log}M,
\end{equation}
where $p(M_{\mathrm{J}})\, \mathrm{d}M_{\mathrm{J}}$ is the distribution of Jeans masses, and the integral over it yields the fraction of cores of mass $M$ that are Jeans unstable. \citet{2002ApJ...576..870P} suggest that the Jeans mass distribution is linked to the turbulent gas density PDF, which is approximately log-normal with a standard deviation given by \citep[see][]{1997MNRAS.288..145P,2008ApJ...688L..79F},
\begin{equation} \label{eq:sigs}
\sigma_{\mathrm{s}}^2=\mathrm{ln}(1+b^2\mathcal{M}^2),
\end{equation}
and therefore, the distribution of Jeans masses is given by \citep{1997MNRAS.288..145P,2002ApJ...576..870P},
\begin{equation}
   P(M_{\mathrm{J}})\, d\mathrm{ln}M = \frac{1}{\sqrt{2\pi}\sigma_{\mathrm{s}}/2} \left(\frac{M_{\mathrm{J}}}{M_{\mathrm{J, 0}}}\right)^{-2} \mathrm{exp}\left[-\frac{1}{2}\left(\frac{\mathrm{ln}M_{\mathrm{J}} - A}{\sigma_{\mathrm{s}}/2}\right)^2\right] d\mathrm{ln}M,
\end{equation}
where $A=\mathrm{ln}M_{\mathrm{J, 0}}^2+\sigma_{\mathrm{s}}^2/2$ and $M_{\mathrm{J, 0}}$ is the mean Jeans mass. For $\beta=2$, which is the typical one-dimensional power spectral index derived for MCs through observations and numerical experiments \citep{2002A&A...390..307O,2004ApJ...615L..45H,2011ApJ...740..120R,2013MNRAS.436.1245F,2021NatAs...5..365F}, the high-mass slope of the IMF based on the \citet{2002ApJ...576..870P} model is
\begin{equation}
    \Gamma=3/(4-\beta)=1.5.
\end{equation}
 The peak of the distribution is then controlled by the scale of the mean thermal Jeans mass $M_{\mathrm{J, 0}}$, which is $\sim 2$--$3\, \mathrm{M_\odot}$ in our simulations.

The \citet{2002ApJ...576..870P} model corresponds to the dash-dotted curves in both the top and bottom panels of Fig.~\ref{fig:imf_theo} and is shown for the simulation input parameters: $\mathcal{M}=5$, $\beta=2$, $M_{\mathrm{J, 0}}=2$, and $b=1$ (COMP) or $b=1/3$ (SOL).

\subsubsection{The \citet{2008ApJ...684..395H} (HC08) model}
 To derive an analytical model for the CMF/IMF, \citet{2008ApJ...684..395H} build upon the framework of the Press-Schechter formalism, which is originally employed in the context of cosmology. Based on the log-normal nature of the gas density PDF for supersonic turbulence, the model associates the self-gravitating structures (analogous to dense cores) with the over-densities in the density distribution that satisfy a collapse criterion. The collapse criterion is defined by the Jeans mass, where the turbulent support is also taken into account. The shape of the derived analytical CMF/IMF is determined by a combination of power-law and log-normal terms. At very small and very large masses, the log-normal term dominates and introduces an exponential cut-off, while the power-law term dominates in the intermediate mass range. The mass scales (both small and large) at which the transition from the power-law to the log-normal form occurs, is determined by the standard deviation of the density PDF ($\sigma_{\mathrm{s}}$), which in turn is dependent on the Mach number and the driving of the turbulence (see Eq.~\ref{eq:sigs}). \citet{2008ApJ...684..395H} argue that the power-law slope is steeper if the non-thermal support against collapse, e.g., the turbulent pressure, is not taken into account. Utilising their expression for the slope of the power-law contribution, which is defined by the turbulence power spectral index $\beta=2$ \citep[][as also assumed in the Padoan \& Nordlund model above]{2021NatAs...5..365F}, \citet{2008ApJ...684..395H} find
\begin{equation}
    \Gamma \approx (\beta+3)/2\beta=1.25.
\end{equation}
We remark that \citet{2013ApJ...770..150H} later incorporated the time dependence of the gas density PDF in their derivation of the CMF/IMF, and found that as a consequence, the power-law slope steepens slightly. We plot the \citet{2008ApJ...684..395H} CMF/IMF by using Eq.~44 in \citet{2008ApJ...684..395H}, again with the simulation input parameters: $\mathcal{M}=5$, $\beta=2$, $M_{\mathrm{J, 0}}=2\, \mathrm{M_\odot}$, $\mathcal{M_*}=1.4$, and $b=1$ (COMP) or $b=1/3$ (SOL). We note that $\mathcal{M_*}$ is the effective sonic Mach number on the scale of the mean Jeans length. The dotted curves in the top and bottom panels of Fig.~\ref{fig:imf_theo} depict the \citet{2008ApJ...684..395H} model.

\subsubsection{The \citet{2012MNRAS.423.2037H} (H12) model}
\citet{2012MNRAS.423.2037H} suggest that in order to accurately derive the mass spectrum of dense cores and subsequently the IMF, the `cloud-in-cloud' problem has to be resolved, i.e., the over-counting arising due to a self-gravitating region being contained in another self-gravitating structure of larger size. Extending the excursion-set formalism to the case of log-normal gas density fluctuations in the ISM, they propose that the mass function of self-gravitating objects on the largest scales (first crossing distribution) represent the mass distribution of giant molecular clouds, while the mass spectrum of self-gravitating objects on the smallest scales (last crossing distribution) corresponds to the CMF/IMF. The absolute mass scale and the dispersion in the gas density PDF are calculated by taking into the account the effects of gas properties on all scales up to the scale of a galactic disc. The relation defining the mass required for collapse at different scales reduces to a Jeans criterion on very small scales and to a Toomre criterion on galactic-disc scales. The Mach number at the driving scale of turbulence or equivalently the Mach number $\mathcal{M}_{h}$ on the galactic scale height significantly influences the shape of the mass function. The \citet{2012MNRAS.423.2037H} mass function has a power-law form in the high-mass regime, which flattens at the turnover mass $M_{\mathrm{sonic}}$, characterised by the sonic scale $R_{\mathrm{sonic}}$, i.e., the scale at which the gas flow becomes subsonic \citep{2021NatAs...5..365F}. We employ the Python code developed by \citet{2021MNRAS.503.1138N} to reproduce the \citet{2012MNRAS.423.2037H} mass function (dashed line in Fig.~\ref{fig:imf_theo}). We mention that here we define $\mathcal{M}_{h}=5$, which is the Mach number representing the velocity dispersion on the driving scale of the turbulence ($L/2$) in the simulations. Due to the periodic nature of the computational domain, our simulations do not have a characteristic scale height. \citet{2021MNRAS.503.1138N} find that such an uncertainty in the distinction of $\mathcal{M}_{h}$ can significantly affect the shape of the IMF as predicted in the \citet{2012MNRAS.423.2037H} model.

\subsubsection{Comparison of the PN02, HC08, and H12 models}
Our SMDs qualitatively agree with the above three theoretical models on the fact that an increase in the relative strength of compressive modes of the turbulence driving results in an increase in the number of low-mass stars formed. In all three turbulence-regulated theories of the IMF, this is because a purely compressive turbulence driving results in a larger standard deviation of the gas density PDF \citep{2008ApJ...688L..79F}, i.e., a higher fraction of high-density gas, which corresponds to a relatively lower Jeans mass.

In the case of COMP (top panel in Fig.~\ref{fig:imf_theo}), the forms of the three gravo-turbulent models agree with each other and compare reasonably well with our SMD, although they slightly underestimate the very low-mass range. In the case of SOL (bottom panel in Fig.~\ref{fig:imf_theo}), the HC08 model, compared to the PN02 and H12 models, matches marginally better with our SMD in the high-mass and low-mass regime. The underestimation of the very low-mass range is more apparent in the case of the SOL model where all the three theoretical models drop off exponentially as they approach the very low-mass regime, sharper than our SMD. \citet{2021MNRAS.507.2448M} also found that these gravo-turbulent models underestimate the very low-mass regime of the simulation SMDs (in that study a natural mixture of turbulence driving modes, $b\sim0.4$, was used). This suggests that the theoretical IMF models principally based on fragmentation promoted by turbulence underestimate the BD population.

\begin{table}
    \centering
	\caption{Comparison of the median and peak masses (at SFE = 5\%) obtained for the simulation SMDs with that of the CMF/IMFs predicted by different theoretical models for the input parameters relevant to our simulations.}
	\label{tab:mpeak}
	\begin{tabular}{l|c|c|c|c|c|c} % four columns, alignment for each
	    \hline
		\hline
		 \multirow{2}{*}{Model} & \multicolumn{5}{c}{Median mass\, $[\mathrm{M_{\odot}}]$}\\ \cline{2-6}
		 & Sim & PN02 & HC08 & H12 & BB05\\
        (1) & (2) & (3) & (4) & (5) & (6)\\
		\hline
		\hline 
		COMP  & $0.37$ & $0.47$ & $0.19$ & $0.47$ & $0.35$\\
		SOL & $0.62$ & $1.54$ & $1.02$ & $5.11$ & $0.61$\\
		\hline
		\hline
		\multirow{2}{*}{Model} & \multicolumn{5}{c}{Peak mass\, $[\mathrm{M_{\odot}}]$}\\ \cline{2-6}
		 & Sim & PN02 & HC08 & H12 & BB05\\
        (1) & (2) & (3) & (4) & (5) & (6)\\
		\hline
		\hline 
		COMP & $0.53$ & $0.44$ & $0.17$ & $0.42$ & $0.47$\\
		SOL  & $0.76$ & $1.39$ & $0.87$ & $5.11$ & 0.82\\
		
		\hline
	\end{tabular}\\
\raggedright\textbf{Notes.} The median mass of the simulation SMDs presented here corresponds to the median sink particle mass and the peak mass presented here for the simulation SMDs corresponds to the peak of the fit to the SMDs obtained using MCMC sampling.
\end{table}

We stress that the gravo-turbulent models discussed here essentially derive the mass distribution of unstable dense cores, analogous to the CMF. Although some observational studies suggest that the shape of the IMF is arguably similar to that of the CMF, the associated mass scales are different \citep{1998A&A...336..150M,1998ApJ...508L..91T,2000ApJ...545..327J,2007MNRAS.374.1413N,2007A&A...462L..17A}. Further, the theoretical models here are compared based on their match with the IMF produced by our simulations, which have limitations in the maximum achievable resolution (see Sec.~\ref{sec:limitations}). The agreement of these theoretical models with the individual-star IMF from observations is a different question which is out of the scope of the present study. The three gravo-turbulent models in Fig.~\ref{fig:imf_theo} have been shifted along the mass-axis so as to fit our SMDs and enable a comparison between their shapes. The median mass and the position of the peak of the three theoretical CMF/IMFs before the mass-shift are compared with the same for the simulation SMDs in Tab.~\ref{tab:mpeak}. While the peak mass of the COMP SMD is lower than that of the SOL SMD by a factor of $\sim1.4$, the peak of the PN02, HC08, and H12 models shifts to lower masses by a factor of 3, 5 and 12, respectively, on changing the input parameter $b$ from $1/3$ (SOL) to $1$ (COMP).

A direct comparison of the models with our SMDs is rational only if a one-to-one mapping between the CMF and IMF can be fully established. It is possible that the IMF may deviate from the CMF due to further fragmentation of the cores, the influence of protostellar outflows, and due to dynamical encounters between the stars, which can terminate accretion. On the other hand, the \citet{2005MNRAS.356.1201B} model represents a different class where the IMF emerges as a result of stars accreting competitively from a common reservoir of gas, until they are dynamically ejected. This is fundamentally distinct from the gravo-turbulent models where the mass of a star is predetermined at the (gas) core level. The \citet{2005MNRAS.356.1201B} model on the other hand derives the IMF from the stellar properties, e.g., the mean accretion rate, and therefore is more directly related to our SMDs (discussed next).

\subsubsection{The \citet{2005MNRAS.356.1201B} (BB05) model}
According to the \citet{2005MNRAS.356.1201B} IMF model, the final mass of a star is controlled by the interplay between accretion and stochastic ejections. All objects, whether stellar or sub-stellar, form with the same mass set by the opacity limit of fragmentation. The objects continue accreting at a constant rate and grow in mass until they are dynamically ejected from the parent cloud, which terminates their accretion. A log-normal function is assumed for the distribution of accretion rates ($P(\dot M_{{\mathrm{acc}}})$), and the probability of an object to be ejected at any given time $e(t)$ is proportional to $\exp(-t/t_{\mathrm{eject}})$, where $t_{\mathrm{eject}}$ is the characteristic ejection timescale. Given the mass of an object $M=M_{\mathrm{min}}+\dot M_{\mathrm{acc}} t$ at time $t$, where $M_{\mathrm{min}}$ is the minimum stellar mass set by the opacity limit of fragmentation and $\dot M_{\mathrm{acc}}$ is the time-averaged accretion rate, the probability distribution for the mass of an object is \citep{2005MNRAS.356.1201B},
\begin{equation}
f(M, t) = \frac{1}{\sqrt{2\pi}\sigma_{\mathrm{acc}}(M - M_{\mathrm{min}})} \mathrm{exp}\left\{-\frac{\left[\mathrm{log}\left(\frac{M - M_{\mathrm{min}}}{t}\right)-\mathrm{log}\overline{\dot M}_{\mathrm{acc}}\right]^2}{2\sigma_{\mathrm{acc}}^2}\right\}.
\label{eq:bate1}
\end{equation}
When the termination of accretion via ejection is taken into account, the mass function becomes \citep{2005MNRAS.356.1201B},
\begin{equation}
f(M) = \int_{0}^{t_{\mathrm{p}}}\,f(M, t)\,e(t)\,dt,
\label{eq:bate2}
\end{equation}
where the time period $t_\mathrm{p}$ corresponds to the time elapsed between the formation of the first star and the end of the simulation. We fit the \citet{2005MNRAS.356.1201B} IMF to our sink mass distribution for the SOL and COMP turbulence driving models by evaluating the following parameters: the mean accretion rate $\overline{\dot M}_{\mathrm{acc}}$, the standard deviation in the accretion rates (in logarithmic units) $\sigma_{\mathrm{acc}}$, the characteristic ejection time $t_{\mathrm{eject}}$, the minimum stellar mass $M_{\mathrm{min}}$, and the time period of the cluster formation $t_\mathrm{p}$. The turnover mass and the width of the IMF are characterised by the quantity $\overline{\dot M}_{\mathrm{acc}}\,t_{\mathrm{eject}}$ and the standard deviation in the accretion rates, respectively. The minimum stellar mass $M_{\mathrm{min}}$ defines the low-mass cut-off of the fit. The parameter values calculated for the SOL and COMP models are listed in Tab.~\ref{tab:bbparams}, which represent the averages of the parameter values derived in the two sets (COMP and SOL). The solid curves in the top and bottom panels of Fig.~\ref{fig:imf_theo} show the \citet{2005MNRAS.356.1201B} IMF model. The median mass and peak of the \citet{2005MNRAS.356.1201B} IMF for the COMP and SOL models are shown in Tab.~\ref{tab:mpeak}. The peak of the \citet{2005MNRAS.356.1201B} IMF for the COMP case is lower than that for the SOL case by a factor of $1.7$.

We see that the \citet{2005MNRAS.356.1201B} model provides a very good fit to both the COMP and SOL SMDs, especially in the sub-stellar regime, which was underestimated by the gravo-turbulent fragmentation models (PN02, HC08, H12). This suggests that it is essential to take into account the dynamical ejections to fully explain the IMF \citep[see also][]{2004MNRAS.347L..47B,2010MNRAS.405..401D,2011ApJ...743...98M,2014MNRAS.439..234M}. The reason why the COMP SMD compares reasonably well with the turbulent fragmentation models in contrast to the SOL SMD is because the COMP simulations have not undergone much time evolution, and therefore the effects of competitive accretion and dynamical ejections that are central to the \citet{2005MNRAS.356.1201B} model, are comparatively low. In fact, in the case of the SOL model, if we consider only the stars that form in the early stages, i.e., those that form in an environment reflecting the initial conditions, then the number of low-mass stars is very small and matches the predictions of the gravo-turbulent models very well (see top left panel in Fig.~\ref{fig:imf_evol}).

Thus, we find that elements of both classes of IMF theoretical models, namely the gravo-turbulent and the competitive accretion/ejection models, are relevant for a comprehensive understanding of the IMF.

\begin{table*}
	\caption{Calculated parameter values for the \citet{2005MNRAS.356.1201B} IMF model.}
	\label{tab:bbparams}
	\begin{tabular}{lccccc} % four columns, alignment for each
	    \hline
		\hline
		 Model & $\overline{\dot M}_{\mathrm{acc}}\, [\mathrm{M_{\odot}}\, \mathrm{yr^{-1}}]$ & $\sigma_{\mathrm{acc}}\, [\mathrm{dex}]$ & $t_{\mathrm{eject}}\, [\mathrm{yr}] $  & $t_\mathrm{p}\, [\mathrm{yr}]$\\
        (1) & (2) & (3) & (4) & (5) \\
		\hline
		\hline 
		COMP  & $1.6 \times 10^{-5}$ & $0.32$ & $3.7 \times 10^4$ & $9.7 \times 10^4$\\
		SOL & $4.9 \times 10^{-6}$ & $0.26$ & $1.9 \times 10^5$ & $7.8 \times 10^5$\\
		
		\hline
	\end{tabular}
	\\
    \raggedright\textbf{Notes.} The values presented here are averages of the parameter values obtained from the multiple simulations (realisations of the turbulence) for each simulation model, COMP and SOL. The \citet{2005MNRAS.356.1201B} IMF fits (solid curves in Fig.~\ref{fig:imf_theo}) have been derived by substituting these parameter values into Eqs.~\ref{eq:bate1}--\ref{eq:bate2} and setting $M_{\mathrm{min}}= 0.01\, \mathrm{M_\odot}$ as the low-mass cut-off of the fit for both the simulation sets.
\end{table*}

% ==========================================
\section{Stellar multiplicity and angular momentum}
\label{sec:Multi}

\subsection{Multiplicity fraction}
We follow the algorithm used in \citet{2009MNRAS.392..590B} to identify the multiple stellar systems in our simulations. We find the closest gravitationally-bound pair (binary) in the list of $N$ individual sink particles that form in a simulation. The closest bound pair is recorded as binary, and then replaced by a single object having the mass, centre-of-mass position and velocity equal to the original bound pair. Now the list consists of $N-2$ single objects and $1$ binary object. In the new list, we search again for the closest pair of bound objects. In case the pair comprises of a binary object and a single object, then they are replaced by a triple. This procedure of replacing the closest bound pair with an object of higher order is carried out repetitively until none of the objects existing in the list are bound to one another or a quintuple is the only feasible outcome of the new pairing. We reject quintuples and systems of higher order, because most high-order multiple systems are dynamically unstable and will most likely decay to lower-order systems with further evolution of the cloud. 

This iterative process transforms a list of individual sink particles into a list of single, binary, triple and quadruple systems, with none being a subset of a system of higher order. For example, none of the objects identified as binaries by the algorithm is a part of a triple or quadruple system. The multiplicity fraction in different mass ranges can be obtained by calculating the ratio of the number of multiple systems to the total number of systems whose primary star lies within the given mass range. Thus, the multiplicity fraction ($mf$) is defined as
\begin{equation} \label{eq:mf}
    mf =  \frac{B + T + Q}{S + B + T + Q},
\end{equation}
where $S, B, T$, and $Q$ denote the number of singles, binaries, triples, and quadruples, respectively, whose primary star mass is within the range for which $mf$ is to be evaluated.

\begin{figure}
    \centering
    \includegraphics[width=\columnwidth]{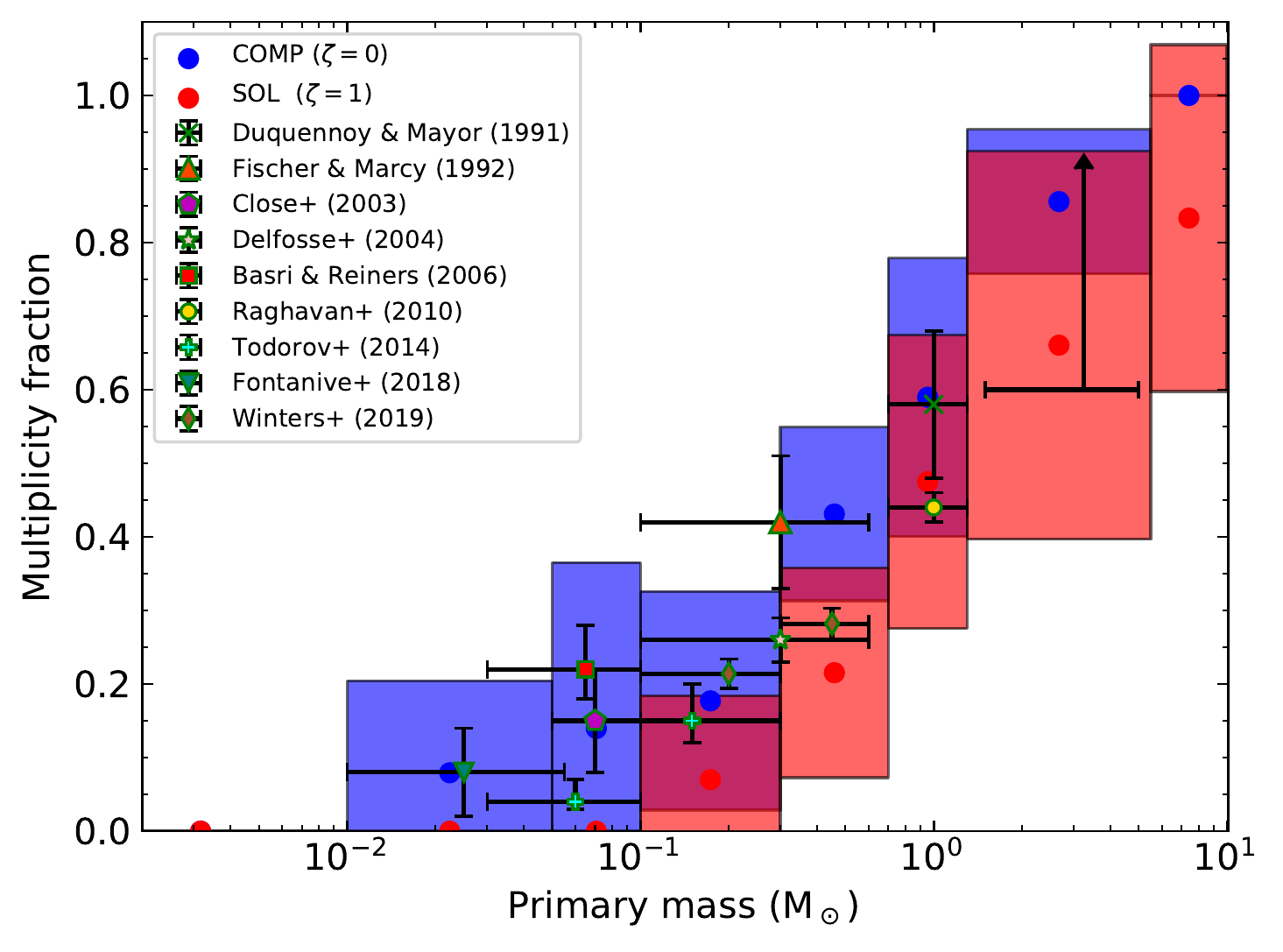}
    \caption{Multiplicity fraction ($mf$) computed via Eq.~(\ref{eq:mf}) in different primary mass intervals for the COMP (blue circular markers and boxes) and SOL (red circular markers and boxes) models. The circular markers denote the average $mf$, obtained across multiple simulations, in the mass interval represented by the width of the patch enclosing the marker. The height of the patch depicts the standard deviation of $mf$ obtained from all the simulations. The centre of the crosses represents the value of $mf$ obtained in different observational studies, with the horizontal and vertical error bars representing the mass range considered in the survey and the uncertainties, respectively. The observational data are (from low to high primary mass), from \citet{2018MNRAS.479.2702F}, \citet{2014ApJ...788...40T}, \citet{2006AJ....132..663B}, \citet{2003ApJ...587..407C}, \citet{2014ApJ...788...40T}, \citet{2019AJ....157..216W} (not corrected for undetected companions), \citet{2004ASPC..318..166D}, \citet{1992ApJ...396..178F}, \citet{2010ApJS..190....1R} and \citet{1991A&A...248..485D}. The multiplicity fraction of high-mass stars is relatively poorly understood. The lower limit of $mf$ in the mass range of $1.5$--$5\,\mathrm{M_\odot}$ is $\sim$ 0.5--0.6 \citep{2012MNRAS.424.1925C,2013ARA&A..51..269D}. Massive stars are expected to have $mf\sim1$ \citep{2009AJ....137.3358M,2011IAUS..272..474S,2017A&A...599L...9S,2020SSRv..216...70L}.}
    \label{fig:Multiplicityfraction}
\end{figure}

\begin{figure}
    \centering
    \includegraphics[width=\columnwidth]{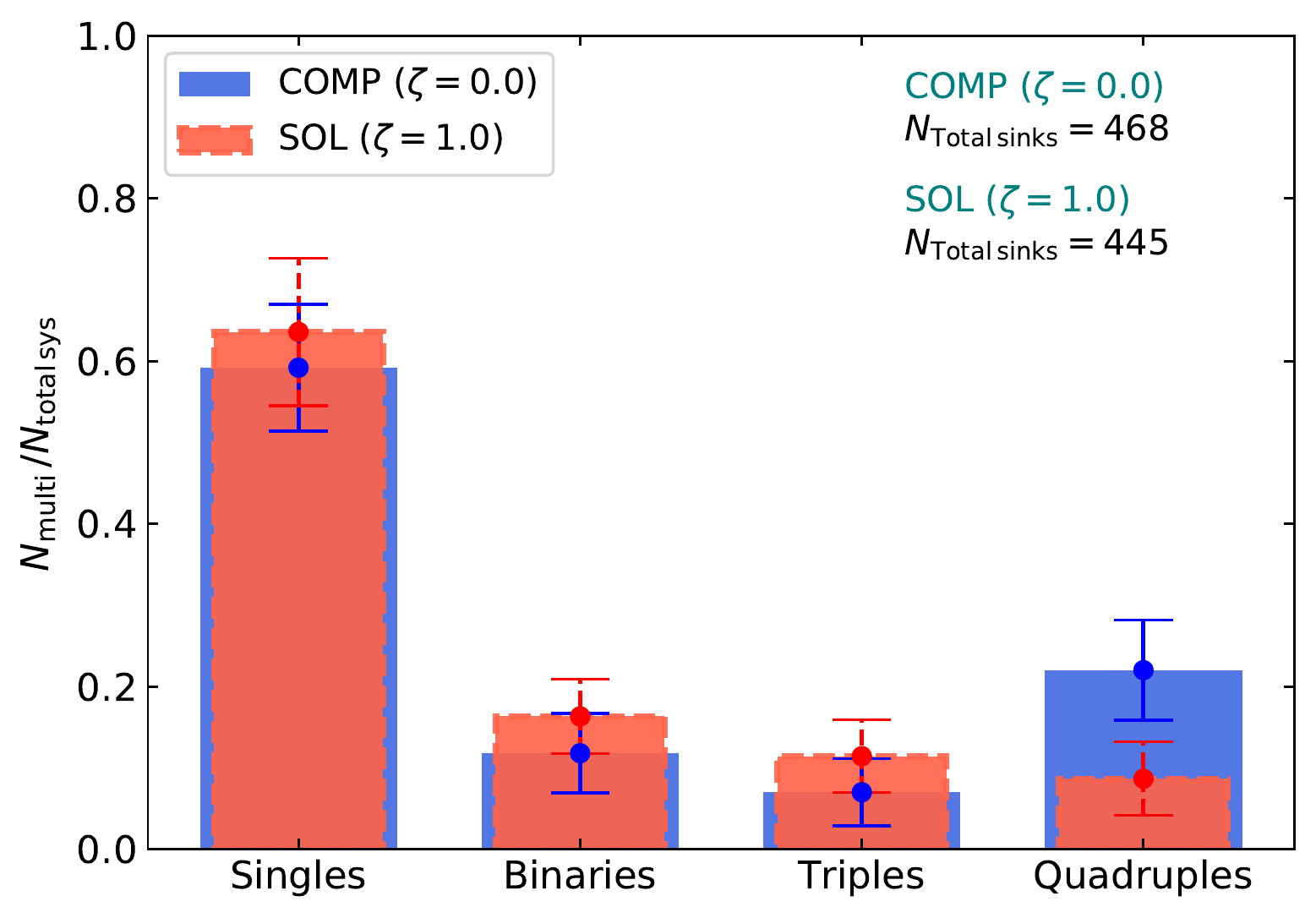}
    \caption{Fraction of single stars and multiple systems (binaries, triples, and quadruples), using the same data as for Fig.~\ref{fig:Multiplicityfraction}. The bars with solid edges correspond to the fractions derived for the COMP model and the bars with dashed edges correspond to the SOL model. The solid and dashed error bars represent the associated standard deviation in the set of simulations for the COMP and SOL models, respectively.}
    \label{fig:ssf_compare}
\end{figure}

Fig.~\ref{fig:Multiplicityfraction} presents the multiplicity fraction as a function of primary mass \citep[also done in][]{2012MNRAS.419.3115B,2012ApJ...754...71K,2018MNRAS.476..771C,2020MNRAS.497..336S,2021MNRAS.507.2448M} at SFE = 5\% for the COMP and SOL models. The mass ranges are selected similar to those chosen in the observational studies so as to allow for a direct comparison. We immediately notice that the multiplicity fraction is an increasing function of the primary mass, for both COMP and SOL, consistent with observational surveys \citep[see the reviews by][]{2013ARA&A..51..269D,2022arXiv220310066O}. However, the multiplicity fraction in each primary mass interval is higher in COMP compared to SOL (see Fig.~\ref{fig:ssf_compare} and the associated text for an explanation).

Our $mf$ values also agree well with observations, except that we are underestimating the multiplicity in the very low-mass stellar (VLMS) and BD ranges. We mention that we do not resolve all of the low-order multiple systems, since the numerical cell width at the highest level of AMR is 100~AU. Therefore, some of the sink particles may actually represent binaries by themselves or triple systems (rarely). However, the numerical resolution effect is expected to be nominal because of the robust nature of the multiplicity fraction definition. The $mf$ value will differ only if a sink particle identified as a single can be further fragmented into multiple individual stars. The value remains unaffected if the sink belongs to a multiple system, i.e., a part of a binary, triple or quadruple object. For example, if a member of a triple system is a binary by itself, then $T\, \mathrm{and}\, Q$ changes to $T-1\, \mathrm{and}\, Q+1$, respectively, which leaves $mf$ unaltered. Based on the observational evidence that the average separation of binaries increases and the frequency of singles decreases with increasing primary mass \citep{2007ApJ...663..394K,2007ApJ...662..413K,2012ARA&A..50...65L}, the mass range that is likely to be affected by the limitation in resolution is the low-mass end, particularly the BD regime. Therefore, the $mf$ values in the sub-solar range (mainly in the regime of M-dwarfs and later types) are expected to be higher than what we obtained for COMP and SOL simulations.

Fig.~\ref{fig:ssf_compare} shows the fraction of singles, binaries, triples and quadruples at SFE = 5\%. The fraction of single stars is highest for both SOL and COMP models, i.e., a large fraction of the sink particles that formed in our simulations are not members of a higher-order multiple system. For the COMP model, the single star fraction (SSF) is $0.59\pm0.08$, while the SSF is $0.64\pm0.09$ for the SOL model (see Tab.~\ref{tab:sims}). While $150$ of the $468$ sink particles formed in the 7~COMP simulations are singles, $177$ of the $445$ sink particles formed in the 11~SOL simulations are singles. Further, the COMP simulations have a substantially higher fraction of quadruples. The total number of quadruples in the COMP simulations is $52$, while there are only $22$ quadruples in the SOL simulations in total. The COMP case is efficient in creating shocked regions of gas that have sufficient mass and high density on average to promote fragmentation into high-order systems, and therefore star formation in these regions is extremely clustered (see Fig.~\ref{fig:turbparam_mapcomp_sfe5}). In addition, the average time period of the COMP simulations is much smaller than that of the SOL simulations (see Tab.~\ref{tab:bbparams}). As a result, the occurrence of dynamic encounters and therefore decay to lower-order systems is low in COMP compared to SOL. This explains the trend of high $mf$ values for the COMP simulations as seen in Fig.~\ref{fig:Multiplicityfraction}. The value of $mf$ is more sensitive to the number of high-order systems. When the number of quadruples increases, the total number of systems (denominator in the $mf$ definition) decreases significantly, leading to high $mf$ values. 

\subsection{Mass ratio distribution}
\begin{figure*}
    \centering
    \includegraphics[width=\textwidth]{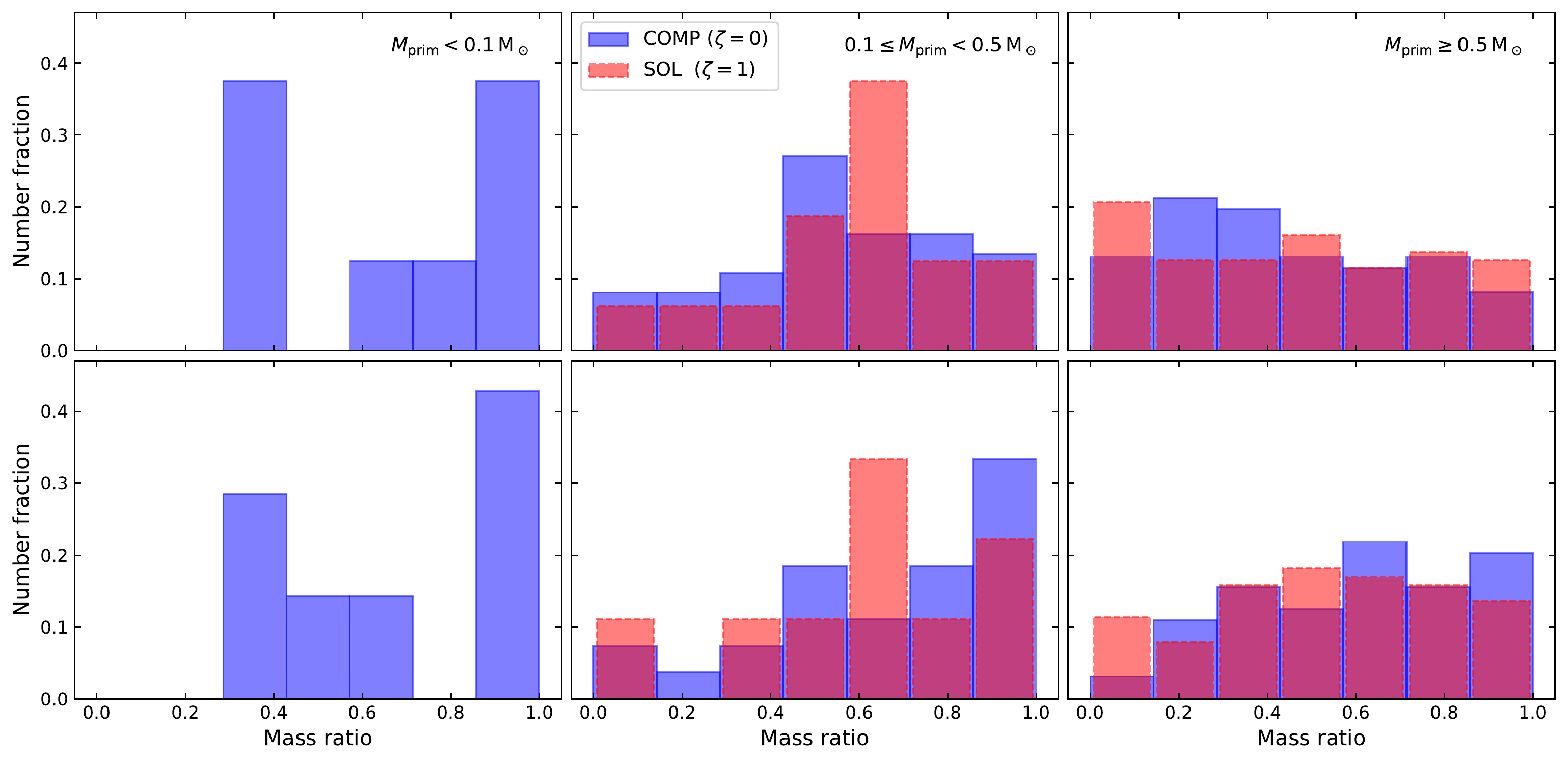}
    \caption{Top panels: Mass ratio distribution of binaries in the multiple systems whose primary mass lies in the range (from left to right) $M_{\mathrm{prim}} < 0.1\, \mathrm{M_\odot}$, $0.1 \leq M_{\mathrm{prim}} < 0.5\, \mathrm{M_\odot}$, and $M_{\mathrm{prim}} \geq 0.5\, \mathrm{M_\odot}$. From each system, the closest binaries are selected for the distribution. The histogram with solid edges represents the distribution for the COMP model and the histogram with dashed edges corresponds to the SOL models. Bottom panels: Similar to the respective panels on the top row, but here, instead of the closest pair, the most massive two members are selected from each system.}
    \label{fig:Massratio}
\end{figure*}

Fig.~\ref{fig:Massratio} shows the mass ratio distribution of binaries for three different primary mass intervals, where the mass ratio is given by $q = M_2/M_1$ and $M_2 < M_1$. For selecting the pairs in each multiple system to be included in the mass-ratio distribution, we use two approaches: 1) the closest pairs (binaries) from each multiple system are selected---every binary, triple, and quadruple system contributes one mass-ratio value, except a quadruple consisting of two binaries orbiting each other which then contributes two mass-ratio values \citep[similar to what is done in][]{2009MNRAS.392..590B}; 2) the two most massive components from each multiple system are included---every binary, triple and quadruple system contributes only one mass-ratio value, including quadruples consisting of two binaries orbiting each other \citep[similar to what is done in][]{2017MNRAS.468.4093G}.

The left, middle and right panels in the top row of Fig.~\ref{fig:Massratio} present the binary mass ratio distributions obtained for the systems whose primary star is in the mass interval $M_{\mathrm{prim}} < 0.1\, \mathrm{M_\odot}$, $0.1 \leq M_{\mathrm{prim}} < 0.5\, \mathrm{M_\odot}$, and $M_{\mathrm{prim}} \geq 0.5\, \mathrm{M_\odot}$, respectively, using the approach similar to that in \citet{2009MNRAS.392..590B}. The bottom row shows the same, but the approach similar to that in \citet{2017MNRAS.468.4093G} is used to obtain the distribution here. In the left panels, there is no mass-ratio distribution for the SOL model because no multiple systems were derived in the primary mass range $M_{\mathrm{prim}} < 0.1\, \mathrm{M_\odot}$ in the case of the purely solenoidal simulations. The mass-ratio distributions obtained using the two approaches seem to be somewhat different, especially in the primary mass range $M_{\mathrm{prim}} \geq 0.5\, \mathrm{M_\odot}$ (right panels). For the primary mass range $M_{\mathrm{prim}} \geq 0.5\, \mathrm{M_\odot}$, the top panel has a slightly higher fraction of pairs with $q < 0.5$ while the bottom has a marginally higher fraction of pairs with $q > 0.5$. Stars with masses in the solar range and higher are generally members of high-order systems, i.e., triples and quadruples. Therefore, the choice in the approach used for selecting the binary pairs for the mass-ratio distribution is expected to affect the distribution.

Overall, irrespective of the turbulence driving mode or the method used for selecting the binaries, the mass ratio distribution for the mass range $M_{\mathrm{prim}} \geq 0.5\, \mathrm{M_\odot}$ is relatively flat, while the distributions for the mass ranges $0.1 \leq M_{\mathrm{prim}} < 0.5\, \mathrm{M_\odot}$ and $M_{\mathrm{prim}} < 0.1\, \mathrm{M_\odot}$ clearly have higher fractions of pairs with $q > 0.5$, which is consistent with the mass-ratio distributions derived from observations in the solar, M-dwarf and VLM regimes, respectively \citep[see reviews by][]{2010ApJS..190....1R,2022arXiv220310066O}. We note that observational surveys find that the mass-ratio distribution is also dependent on the orbital period or separation of the binary \citep[e.g., ][]{1997AJ....113.2246R,2011AJ....141...52T,2012AJ....144...64D,2015MNRAS.449.2618W,2017ApJS..230...15M}. Here, we do not make such a distinction while producing the mass-ratio distributions.

\subsection{Specific angular momentum of dense cores and stars}
\label{sec:angmom}
The evolution of angular momentum from the early stages of the collapse of a dense core to the formation of a main sequence star is a highly debated topic. The specific angular momentum ($j$) of dense molecular cloud cores (diameter $\sim\!0.1\,$pc) is found to be greater than $10^{21}\, \mathrm{cm^2\, s^{-1}}$ \citep{1993ApJ...406..528G,2000ApJ...543..822B,2002ApJ...572..238C}. The specific angular momentum regime of class~0/I envelopes and binary systems is $10^{17}$--$10^{21}\, \mathrm{cm^2\, s^{-1}}$ \citep{1992ASPC...32...41S,1997ApJ...488..317O,2015ApJ...799..193Y}, while that of T-Tauri stars is $10^{16}$--$10^{17}\, \mathrm{cm^2\, s^{-1}}$ \citep{1986ApJ...309..275H}. \citet{2020A&A...637A..92G} find that the $j$ value of class~0 protostellar envelopes is virtually constant, at around $10^{20}\, \mathrm{cm^2\, s^{-1}}$, from a scale of $\sim\!$~1600~AU to 50~AU.

\begin{figure}
    \centering
    \includegraphics[width=\columnwidth]{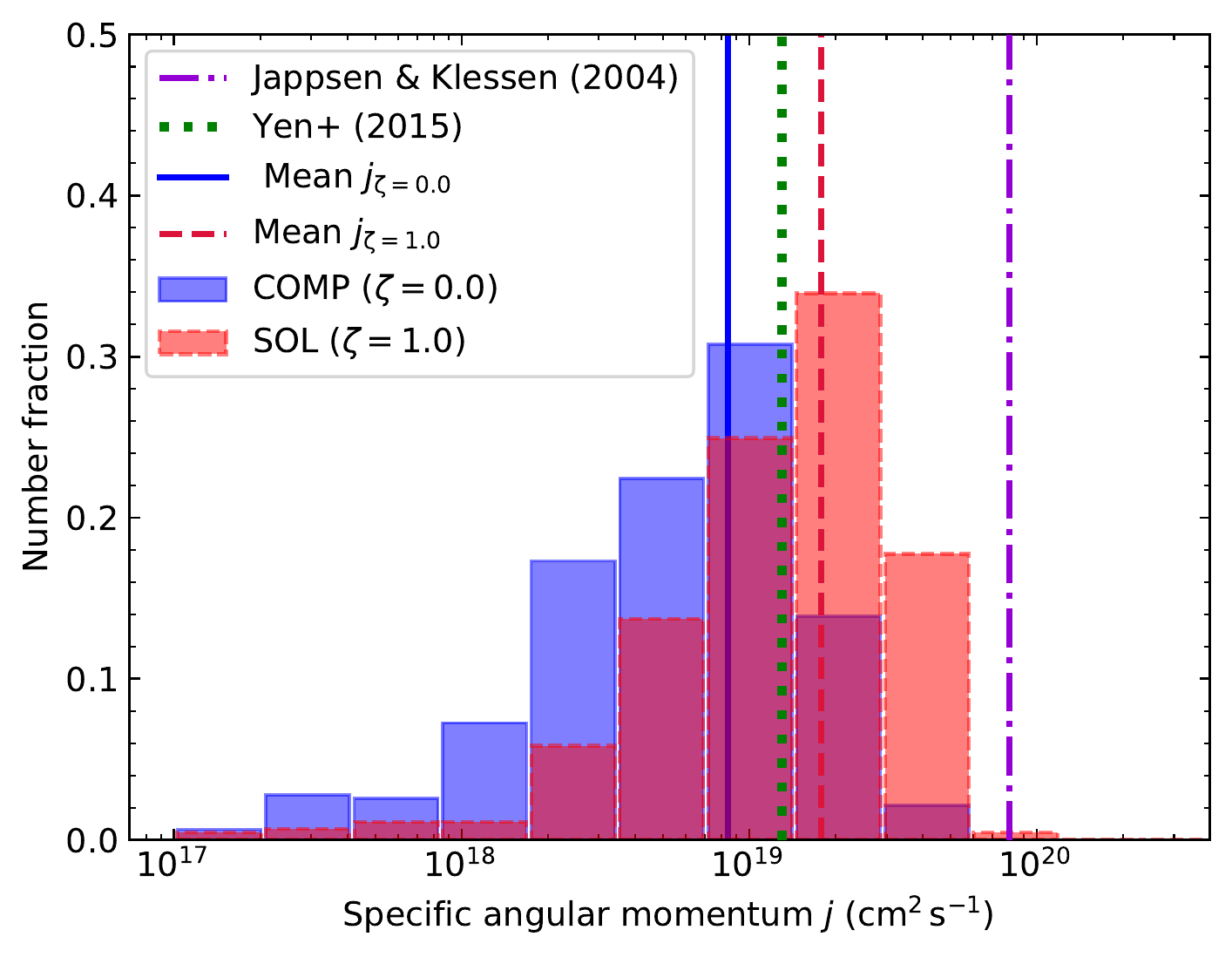}
    \caption{Specific angular momentum $j$ of the sink particles ($r_{\mathrm{sink}}= 250\, \mathrm{AU}$) from the COMP (histogram with solid edges) and SOL (histogram with dashed edges) simulations. The solid and dashed lines correspond to the mean $j$ value of the COMP and SOL models. The dotted line presents the $j$ value measured for the class~0 protostar B335 at $\sim\!$~180~AU by \citet{2015ApJ...812..129Y}, and the dash-dotted line represents the mean value of $j$ obtained in the hydrodynamic simulations of \citet{2004A&A...423....1J} where the sink particle radius is 560~AU.}
    \label{fig:specific_angm}
\end{figure}

\citet{2004A&A...423....1J} carried out hydrodynamic simulations of the collapse of supersonic turbulent clouds and determined $j_{\mathrm{mean}} = 8 \times 10^{19}\, \mathrm{cm^2\, s^{-1}}$ for their sink particles, which have an accretion radius of 560~AU. The specific angular momentum distribution of every sink particle that formed in the simulations of the COMP and SOL models, respectively, is shown in Fig.~\ref{fig:specific_angm}. The range of specific angular momentum of the sinks (having an accretion radius of 250~AU) in both the simulations ($\sim\!10^{17}$--$10^{20}\, \mathrm{cm^2\, s^{-1}}$) spans the regime of protostellar envelopes and binaries, although a small fraction of the sink particles have $j$ values typical of T-Tauri stars. The average specific angular momentum in the COMP model is $j_{\mathrm{mean}} = 8.4 \times 10^{18}\, \mathrm{cm^2\, s^{-1}}$, while that in the SOL model is $j_{\mathrm{mean}} = 1.8 \times 10^{19}\, \mathrm{cm^2\, s^{-1}}$, i.e., about a factor of 2 higher in SOL vs.~COMP. This is most likely because the sink particles in SOL form from gas with $\sim2$ times higher fraction of solenoidal (rotational) modes compared to COMP \citep[see the bottom panel of Fig.~3 in][for $\mathcal{M}=5$]{2011PhRvL.107k4504F}. The $j$ value inferred by \citet{2015ApJ...812..129Y} of the class~0 protostar B335 ($1.3 \times 10^{19}\, \mathrm{cm^2\, s^{-1}}$) measured at a scale of $\sim\!$~180~AU, lies between the average values of the COMP and SOL simulations. 

\section{Discussion}
\label{sec:discussions}

\subsection{The mode of turbulence driving}
 \citet{2010A&A...516A..25S} studied the effect of turbulence driving on the mass distribution of dense cores in simulations where the cores were identified using a clump-finding algorithm. They find that a purely compressive turbulence driving results in a higher fraction of low-mass cores compared to purely solenoidal driving, which qualitatively agrees with our conclusions. Self-gravity is absent in the simulations of \citet{2010A&A...516A..25S}. Consequently, some of the dense cores identified may not be gravitationally bound, and also conversely, some regions that are not identified as bound might become (or have become) gravitationally bound if gravity had been included in their simulations. Moreover, the study assumes that the obtained CMF can be directly mapped to the IMF, which might not be the case in reality \citep{Smith_2009}.
 
 \citet{2015MNRAS.449..662L} carried out multiple simulations of cloud-collapse on the scales of prestellar cores with different turbulence realisations (no driving though) and analysed the dependence of the stellar mass on the variation of the fraction of solenoidal turbulent energy $\mathrm{\delta_{sol}}$. They find that the median stellar mass decreases with increasing $\mathrm{\delta_{sol}}$, contrary to our findings. \citet{2015MNRAS.449..662L} mention that in their simulations with high $\mathrm{\delta_{sol}}$, disc fragmentation dominates over filament fragmentation. Since discs are more prone to multiple fragmentation, the mean stellar mass would be lower in their simulations with high $\mathrm{\delta_{sol}}$, which generate sufficient angular momentum to form discs. However, it is difficult to directly compare their results with that of our relatively large-scale simulations, because, although the large-scale driving is purely solenoidal in our SOL simulations, it is not necessary that the solenoidal modes would always dominate on the scales where stars form, i.e., on the scales of prestellar cores.
 
 Our results also tend to disagree with \citet{2017MNRAS.465..105L} who observed that the simulation IMF obtained from a purely compressive inital velocity field was indistinguishable from the one obtained with a purely solenoidal initial velocity field. However, as in \citet{2015MNRAS.449..662L}, the turbulence was not continuously driven. In such a setup, the turbulence decays rapidly, and therefore the differences between solenoidal and compressive modes is relatively small once the stars begin to form. Another problem with that type of setup is that the initial density field \citep[usually chosen to be uniform or of some radial power-law form,][]{2011MNRAS.413.2741G} is inconsistent with the initial turbulent velocity field; that is, it takes about 2~turbulent cloud crossing times for the turbulence to become fully developed and the density and velocity field to establish reasonable turbulent statistics \citep{2009ApJ...692..364F,2009A&A...508..541K,2010MNRAS.406.1659P}, a time by which the star formation experiments in \citet{2015MNRAS.449..662L} and \citet{2017MNRAS.465..105L} are already completed, and therefore, the turbulence in their simulations is never actually fully developed. This can only be achieved with continuous driving \citep{1998ApJ...508L..99S,1998PhRvL..80.2754M}.
 
 \subsection{The velocity power spectrum}
 \citet{2009MNRAS.397..232B} examined the dependence of the IMF on the kinetic power spectrum of the turbulent gas by comparing cloud-collapse simulations that are initialised with a power spectrum given by $P(k) = k^{-4}$ with those that start with a power spectrum consistent with the Larson scaling relations, i.e., $P(k) = k^{-2}$. They find that the IMFs produced by the two models are nearly indistinguishable. On the other hand, \citet{2021MNRAS.503.1138N} performed a set of simulations driven with a power spectrum given by $P(k) = k^{-2}$ and another set of simulations with the same initial conditions, except that they change the power spectrum of driving to $P(k) = k^{-1}$. They show that the shallower power spectrum results in a shallower high-mass slope in the IMF. The differences in how the turbulence was injected is likely the reason why the results of \citet{2009MNRAS.397..232B} and \citet{2021MNRAS.503.1138N} are contradictory to each other---only an initial turbulent velocity field was imposed in the simulations of \citet{2009MNRAS.397..232B}, while the turbulence in the \citet{2021MNRAS.503.1138N} simulations was driven continuously \citep[see also][]{2022arXiv220510413G}.
 
 It is clear that the conclusions of the above studies vary in terms of the effect of turbulence on the IMF. This is mainly due to the differences in the numerical setup chosen to conduct the experiment, particularly how turbulent motions are introduced in the simulations, e.g., impulsive initial velocity field versus continuous driving, with the former having only limited predictive power (c.f., discussion in the preceding subsection).

 \subsection{Numerical resolution and physics included}
 \label{sec:limitations}
 Another important aspect is the maximum numerical resolution that can be attained in simulations of star formation. Numerical studies like \citet{2009MNRAS.397..232B} and \citet{2017MNRAS.465..105L} can resolve down to very small scales. However, with the increase in resolution, it also becomes important to include more physical mechanisms, such as magnetic fields, stellar heating and mechanical feedback (jets/outflows), which we do. While our simulations cannot capture fragmentation on the smallest scales ($\lesssim100\,\mathrm{AU}$), the resolution is sufficient to compare our simulation SMDs with the system IMFs (unresolved close binaries) from observations. While the limitations in numerical resolution only allow us to compare system IMFs, the inclusion of the aforementioned physics in our simulations is crucial for a comprehensive understanding of the IMF.

\section{Conclusions}
\label{sec:conclude}
We carried out a series of simulations of star cluster formation in molecular clouds incorporating gravity, turbulence, magnetic fields, stellar radiative heating and protostellar outflows to study the influence of the turbulence driving mode on the IMF. We find that the IMF derived for simulations driven by purely compressive modes has a higher fraction of low-mass stars and has a lower characteristic mass (median) as compared to the IMF obtained for simulations driven by purely solenoidal modes. We performed a Kolomogrov-Smirnov test to dismiss the possibility that the differences in the distributions are insignificant. In addition, to quantitatively confirm that the shape of the two distributions differs, we fit a modified version (to account for the finite mass in our numerical domain) of the \citet{2005ASSL..327...41C} IMF, where the parameters including the peak mass, the standard deviation of the log-normal part, the transition mass and the slope of the power-law part of the IMF, are estimated using Markov Chain Monte-Carlo sampling. We find that the IMF parameter sets obtained for purely compressive and purely solenoidal driving primarily differ in the median (characteristic) mass of the IMF, i.e., the IMF from compressive driving is shifted to lower masses by a factor of $\sim1.5$ compared to solenoidal driving.

We find that our simulation SMDs generally agree with the functional form of the IMF derived from different observational studies, i.e., the existence of a power-law tail at high masses and flattening at sub-solar masses. We see that, while the peak of the SMD produced by simulations with purely compressive driving ($\sim 0.3-0.5\, \mathrm{M_\odot}$) is comparable to the peak of the observed IMF ($\sim 0.3\, \mathrm{M_\odot}$), the peak of the SMD corresponding to the purely solenoidally driven simulations is too high ($\sim 0.6-1.0\, \mathrm{M_\odot}$). We also compare our IMFs with various theoretical models of the IMF based on gravo-turbulent fragmentation. We find that the gravo-turbulent models of the IMF \citep{2002ApJ...576..870P,2008ApJ...684..395H,2012MNRAS.423.2037H} successfully predict a decrease in the fraction of low-mass stars on switching from a purely compressive to purely solenoidal turbulence driving, as observed in our simulations. However, the gravo-turbulent models underestimate the number of low-mass stars formed in the purely solenoidal driving simulations, especially in the very low-mass regime. While many stars in the mass range of M dwarfs and later types form in the early stages of the purely compressive driving simulations, the number of such stars that form early is significantly lower in the case of the purely solenoidal driving simulations. A substantial fraction of the low-mass stars in the solenoidal simulations form towards the later stages of the cloud evolution. This explains why the models based on turbulent fragmentation underestimate the low-mass stars in the solenoidal simulations---these models are based on the cloud properties characteristic of a cloud in the early stages of the evolution. The gravo-turbulent models do not consider the time evolution of the parent cloud and stars while they are forming, such as changes in the gas density PDF, fragmentation of discs, and ejections via encounters \citep{2002MNRAS.332L..65B,2007A&A...466..943G,2007MNRAS.382L..30S,2011ApJ...730...32S,2011MNRAS.416.1436B,2012MNRAS.423.1896R,2015ApJ...800...72T,2015ApJ...808...48B,2018A&A...611A..88L,2021MNRAS.507.4335K}. 

We find that our simulation SMDs compare well with the \citet{2005MNRAS.356.1201B} IMF model, which is based on accretion and stochastic ejections of stars. The agreement is most significant in the very low-mass range of the IMF compared with the gravo-turbulent models, emphasising the relevance of dynamical ejections during the formation of sub-stellar objects. However, the \citet{2005MNRAS.356.1201B} IMF model is based on stellar properties, i.e., the mean and dispersion of the accretion rate, and the ejection timescale, as opposed to the gravo-turbulent models, which rely only on gas properties. The \citet{2005MNRAS.356.1201B} model does not address how the IMF depends on stellar feedback and/or the properties of the MHD turbulence in the parental gas cloud, i.e.,  it cannot explain why our array of simulations shows that feedback and MHD turbulence, specifically the mode of driving, plays an important role in setting the characteristic stellar mass and the power-law slope \citep[see also][]{2021MNRAS.503.1138N}. On the other hand, the gravo-turbulent models attempt to predict the shape of the IMF based on the turbulent gas properties only, without taking into the account the important dynamical evolution of the young stars when they interact in dense multiple systems. Therefore, our results and discussion here suggests that the theoretical models of the IMF need further revision, such that both the gas properties of the parental cloud and the dynamical interaction of young stars are self-consistently taken into account.

Our results further suggest that the top-heavy nature of the IMF observed in clouds near the Galactic centre (in the Central Molecular Zone, CMZ) may be (at least partly) a consequence of the turbulence driving properties in those regions---turbulent motions in the CMZ are likely driven by solenoidal modes, as a result of enhanced shear \citep{2016ApJ...832..143F, 2022arXiv220613442R}. As our simulations show, solenoidally-driving turbulence leads to less fragmentation and produces a higher median mass of stars than compressive driving. Therefore, in addition to the increased temperature, a predominately solenoidal driving mode of turbulence in the CMZ may explain the increased Jeans mass and consequently the observed top-heavy IMF in clouds near the Galactic centre \citep[see also][]{2007MNRAS.374L..29K}. The stars that are able to form in these conditions can grow to relatively higher masses, and as a consequence, the increased radiative heating by these stars hinders fragmentation in the later stages of the cloud collapse. Thus, the end result would be a higher fraction of high-mass stars in the CMZ compared to solar-neighbourhood clouds.

Finally, we compare the multiplicity properties of stars formed in purely compressive and purely solenoidal driving simulations. We find that purely compressive driving produces a higher fraction of multiple systems compared to solenoidal driving. For both driving modes, we observe that the multiplicity fraction is a monotonically increasing function of the primary mass, which is consistent with observations. However, compressive driving leads to a relatively higher multiplicity fraction for any primary mass. We find that the mass-ratio distribution of binaries in our simulations agree with observations, and this distribution does not seem to be influenced by the turbulence driving mode. The specific angular momentum $j$ of the sink particles (having an accretion radius of 250~AU) for both solenoidal and compressive driving compares well with the $j$ value obtained for protostellar envelopes and binaries in observational surveys. The mean $j$ value for solenoidal driving is about twice as large as that for compressive driving, as a consequence of the factor $\sim2$ higher solenoidal kinetic energy fraction for solenoidal driving compared to compressive driving.

\section*{Acknowledgements}
We thank the anonymous reviewer for their comments, which helped to improve the paper. C.F.~acknowledges funding provided by the Australian Research Council (Future Fellowship FT180100495), and the Australia-Germany Joint Research Cooperation Scheme (UA-DAAD). We further acknowledge high-performance computing resources provided by the Leibniz Rechenzentrum and the Gauss Centre for Supercomputing (grants~pr32lo, pr48pi and GCS Large-scale project~10391), the Australian National Computational Infrastructure (grant~ek9) in the framework of the National Computational Merit Allocation Scheme and the ANU Merit Allocation Scheme. The simulation software FLASH was in part developed by the DOE-supported Flash Center for Computational Science at the University of Chicago.

%%%%%%%%%%%%%%%%%%%%%%%%%%%%%%%%%%%%%%%%%%%%%%%%%%
\section*{Data Availability}
The data used in this article is available upon reasonable request to the authors.



\bibliographystyle{mnras}
\bibliography{Bibliography} % if your bibtex file is called example.bib

%\begin{thebibliography}{99}
%\end{thebibliography}

%%%%%%%%%%%%%%%%% APPENDICES %%%%%%%%%%%%%%%%%%%%%

\appendix

\section{IMF fit using MCMC sampling}
\label{sec:appendix}
\begin{figure*}
    \centering
    \includegraphics[width=\textwidth]{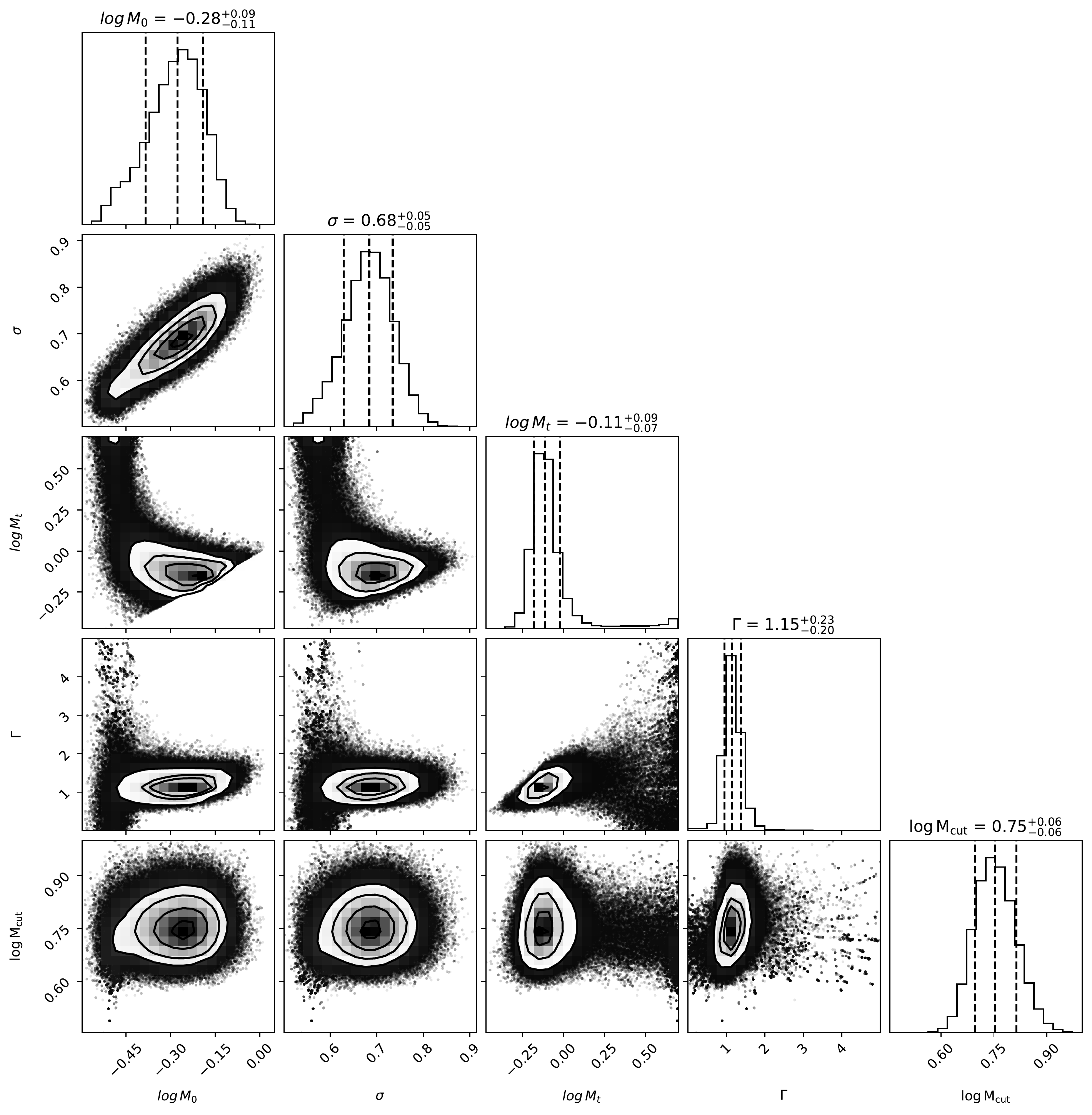}
    \caption{Posterior probability distribution of the parameters corresponding to the IMF fit for the COMP model obtained using MCMC sampling.}  
    \label{fig:cornerplot_blue_withmaxmass}
\end{figure*}

\begin{figure*}
    \centering
    \includegraphics[width=\textwidth]{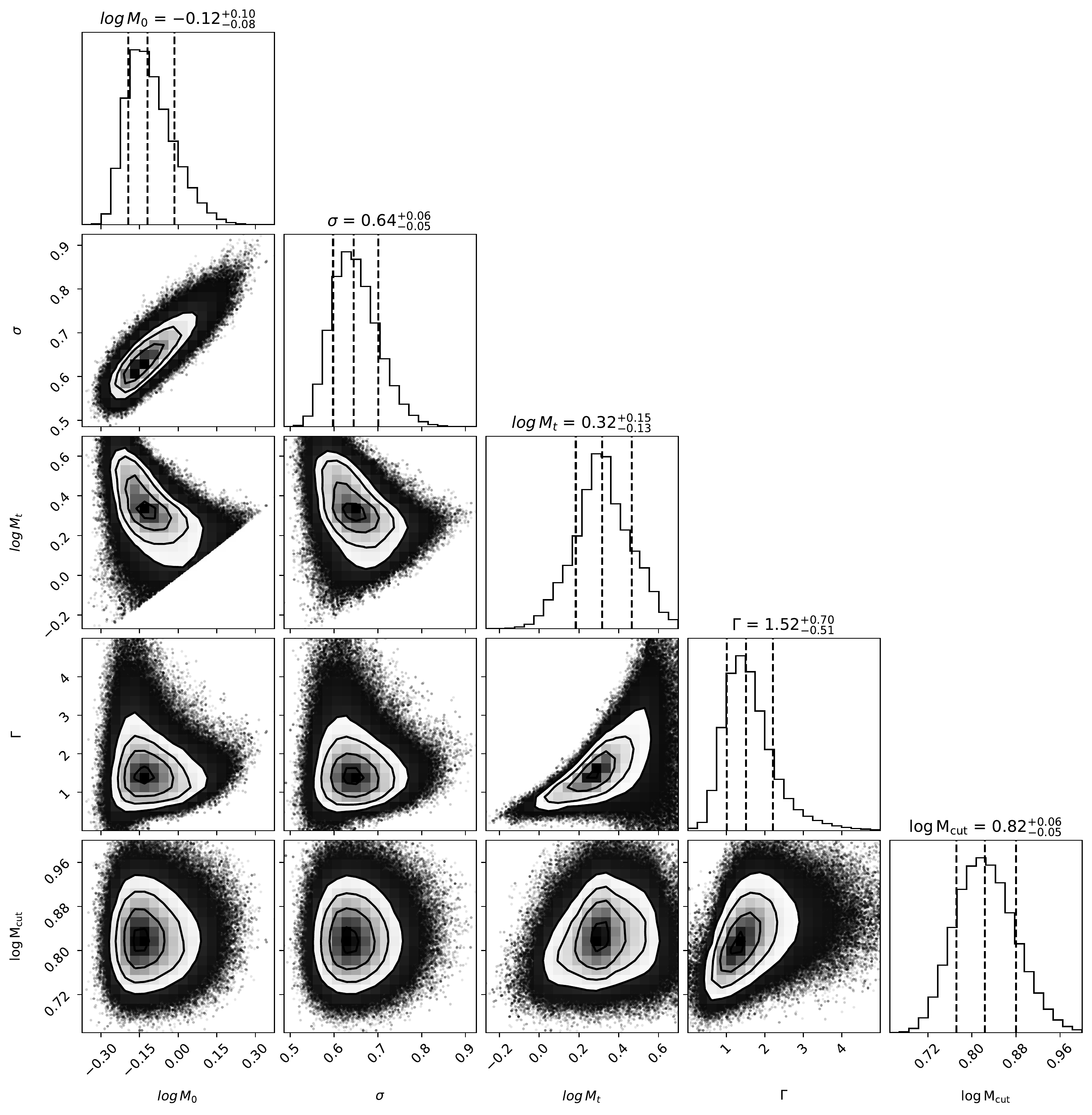}
    \caption{Same as Fig.~\ref{fig:cornerplot_blue_withmaxmass}, but for the SOL model.}  
    \label{fig:cornerplot_red_withmaxmass}
\end{figure*}

\begin{figure}
    \centering
    \includegraphics[width=\columnwidth]{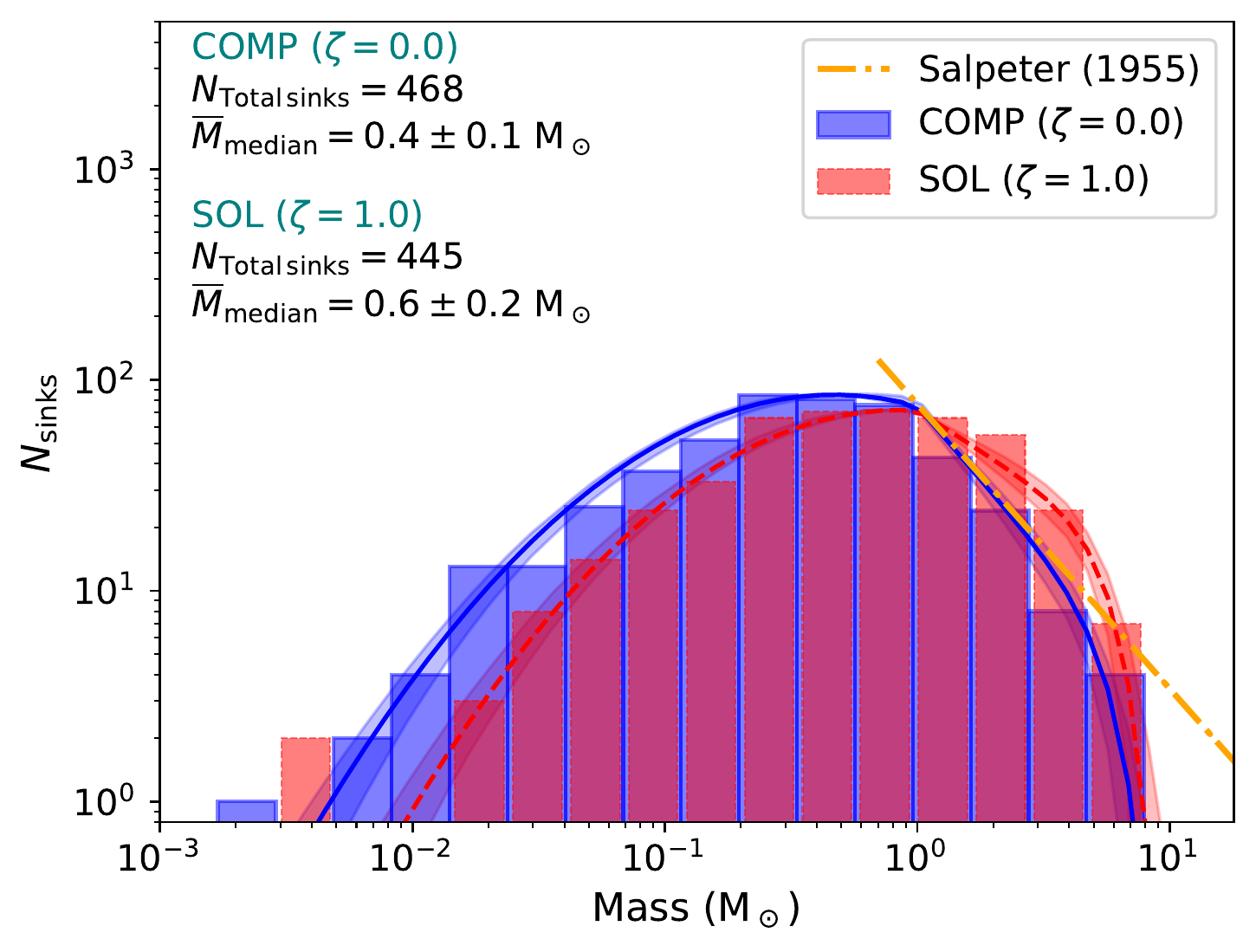}
    \caption{Same as Fig.~\ref{fig:IMF_both}, but the fitted curves (solid and dashed) are based on the parameter values for the fixed $M_{\mathrm{T}}$ case in Tab.~\ref{tab:mcmc}.
    }  
    \label{fig:imf_both_mt_fixed}
\end{figure}

\begin{figure*}
    \centering
    \includegraphics[width=\textwidth]{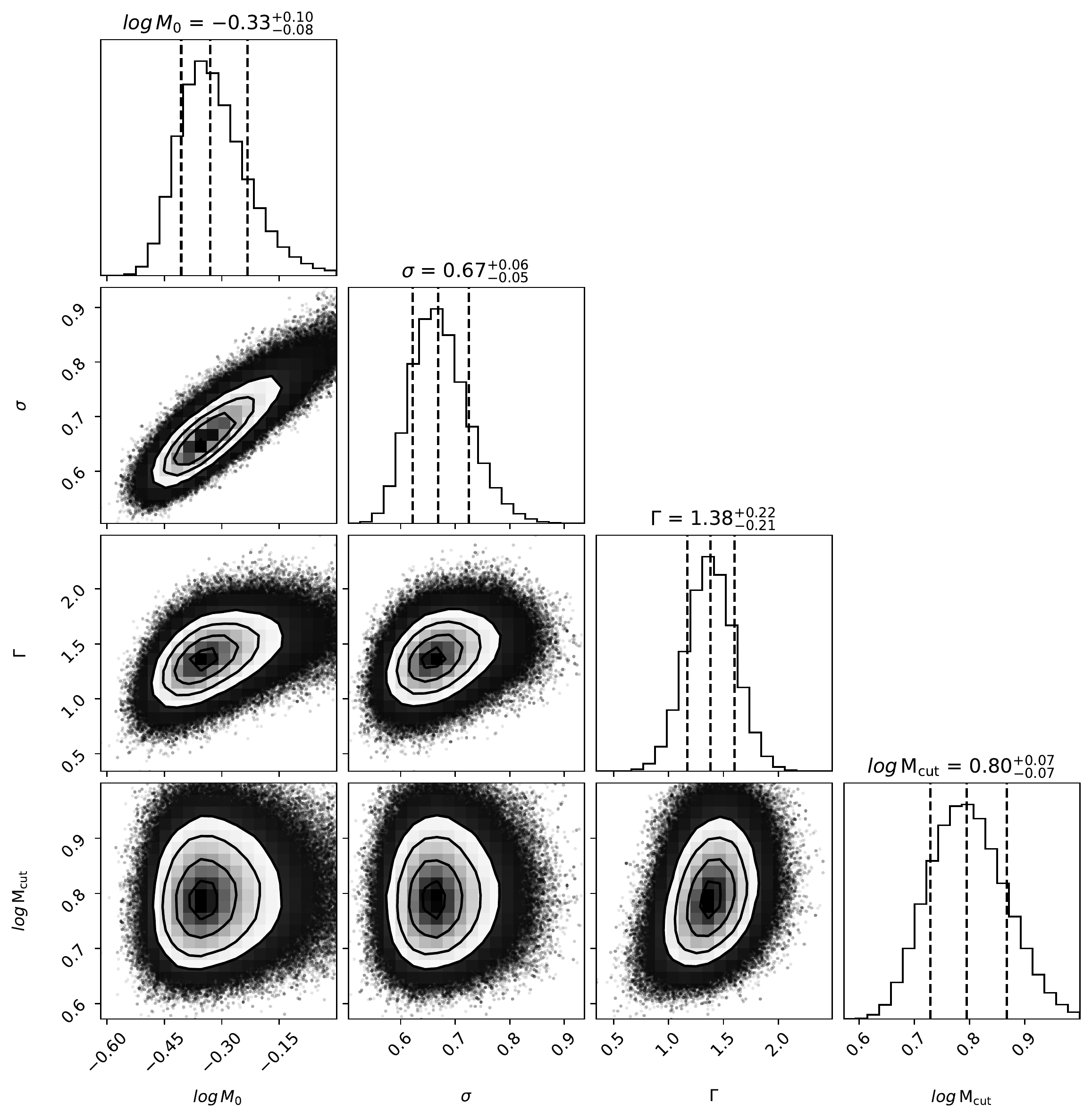}
    \caption{Same as Fig.~\ref{fig:cornerplot_blue_withmaxmass}, but where $M_{\mathrm{T}}$ is fixed.}  
    \label{fig:cornerplot_blue_mt_fixed}
\end{figure*}

\begin{figure*}
    \centering
    \includegraphics[width=\textwidth]{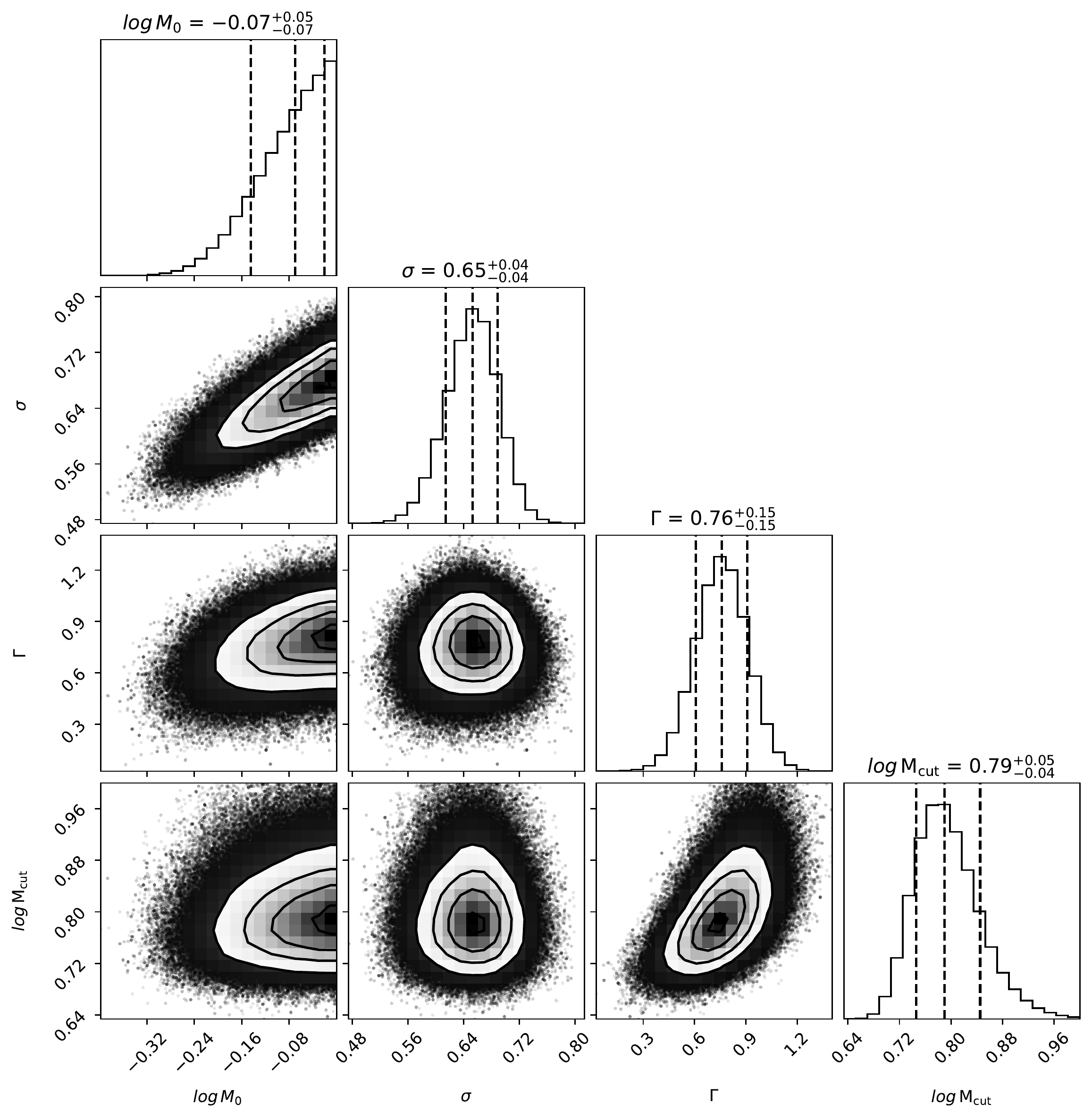}
    \caption{Same as Fig.~\ref{fig:cornerplot_red_withmaxmass}, but where $M_{\mathrm{T}}$ is fixed.}  
    \label{fig:cornerplot_red_mt_fixed}
\end{figure*}

Fig.~\ref{fig:cornerplot_blue_withmaxmass} depicts the corner plots showing the one-dimensional and two-dimensional posterior probability distributions for the different parameters of the IMF fit derived using the MCMC sampling method (free $M_{\mathrm{T}}$ case in Tab.~\ref{tab:mcmc}) in the case of the COMP model, and Fig.~\ref{fig:cornerplot_red_withmaxmass} shows the same for the SOL model. Fig.~\ref{fig:imf_both_mt_fixed} presents the fits obtained for the COMP and SOL SMDs using the values from the fixed $M_{\mathrm{T}}$ case in Tab.~\ref{tab:mcmc}. The corresponding corner plots are shown in Fig.~\ref{fig:cornerplot_blue_mt_fixed} (COMP) and Fig.~\ref{fig:cornerplot_red_mt_fixed} (SOL), respectively. We note that we have added the additional constraint that $M_0 < M_{\mathrm{T}}$, which is the reason for the abrupt cut-off in the posterior distribution of $M_0$ in Fig.~\ref{fig:cornerplot_red_mt_fixed}. We can see that on fixing $M_{\mathrm{T}}=1$, the value of $\Gamma$ changes significantly between SOL and COMP. However, the important feature is that the parameter set associated with the COMP and SOL SMDs are still different.

%%%%%%%%%%%%%%%%%%%%%%%%%%%%%%%%%%%%%%%%%%%%%%%%%%

% Don't change these lines
\bsp	% typesetting comment
\label{lastpage}
\end{document}